\documentclass[pageno]{jpaper}

\usepackage{mathptmx} 
\usepackage{amsmath,amssymb,amsfonts}
\usepackage{graphicx}
\usepackage{textcomp}
\usepackage{xcolor}
\usepackage{color,soul}
\usepackage{booktabs}
\usepackage{tabularx}
\usepackage{balance}
\usepackage{amsbsy}
\usepackage{pifont}
\usepackage{listings}
\usepackage{xspace}
\usepackage[normalem]{ulem}

\usepackage{cleveref}
\crefformat{chapter}{\S#2#1#3} 
\crefformat{section}{\S#2#1#3} 
\crefformat{subsection}{\S#2#1#3}
\crefformat{subsubsection}{\S#2#1#3}

\definecolor{darkred}{rgb}{0.8, 0.0, 0.0}
\definecolor{darkgreen}{rgb}{0.0, 0.5, 0.0}
\definecolor{mustard}{rgb}{1.0, 0.77, 0.05}
\newcommand{\cmark}{\ding{51}}%
\newcommand{\xmark}{\ding{55}}%

\newcommand{\proj}{Tascade\xspace}
\newcommand{\device}{TCA\xspace}

\title{Massive Data-Centric Parallelism in the Chiplet Era}

\author{
Marcelo Orenes-Vera, Esin Tureci, David Wentzlaff, Margaret Martonosi\\
Princeton University, Princeton, New Jersey, USA \\
\{movera, esin.tureci, wentzlaf, mrm\} @princeton.edu
}

\begin{document}
\date{}
\maketitle
\pagestyle{plain}

\newcommand{\subf}[2]{%
  {\small\begin{tabular}[t]{@{}c@{}}{\tiny}
  #1\\#2
  \end{tabular}}%
}

\begin{abstract}


Recent works have introduced task-based parallelization schemes to accelerate graph search and sparse data-structure traversal, where some solutions scale up to thousands of processing units (PUs) on a single chip.
However, parallelizing these memory-intensive workloads across millions of cores requires a scalable communication scheme as well as designing a cost-efficient computing node that makes multi-node systems practical, which has not been addressed in previous research.

To address these challenges, we propose a \textul{ta}sk-oriented \textul{s}calable \textul{c}hiplet \textul{a}rchitecture for \textul{d}istributed \textul{e}xecution (Tascade), a multi-node system design that we evaluate with up to 256 distributed chips---over a million PUs. 
We introduce an execution model that scales to this level via \emph{proxy regions} and \emph{selective cascading}, which reduce overall communication and improve load balancing.
In addition, package-time reconfiguration of our chiplet-based design enables creating chip products optimized post-silicon for different target metrics, such as time-to-solution, energy, or cost.

We evaluate six applications and four datasets, with several configurations and memory technologies, to provide a detailed analysis of the performance, power, and cost of data-centric execution at a massive scale.
Our parallelization of Breadth-First-Search with RMAT-26 across a million PUs---the largest of the literature---reaches 3021 GTEPS.
\end{abstract}




\section{Introduction}

\begin{figure}[t]
\centering  
\vspace{-3mm}
\includegraphics[width=\columnwidth]{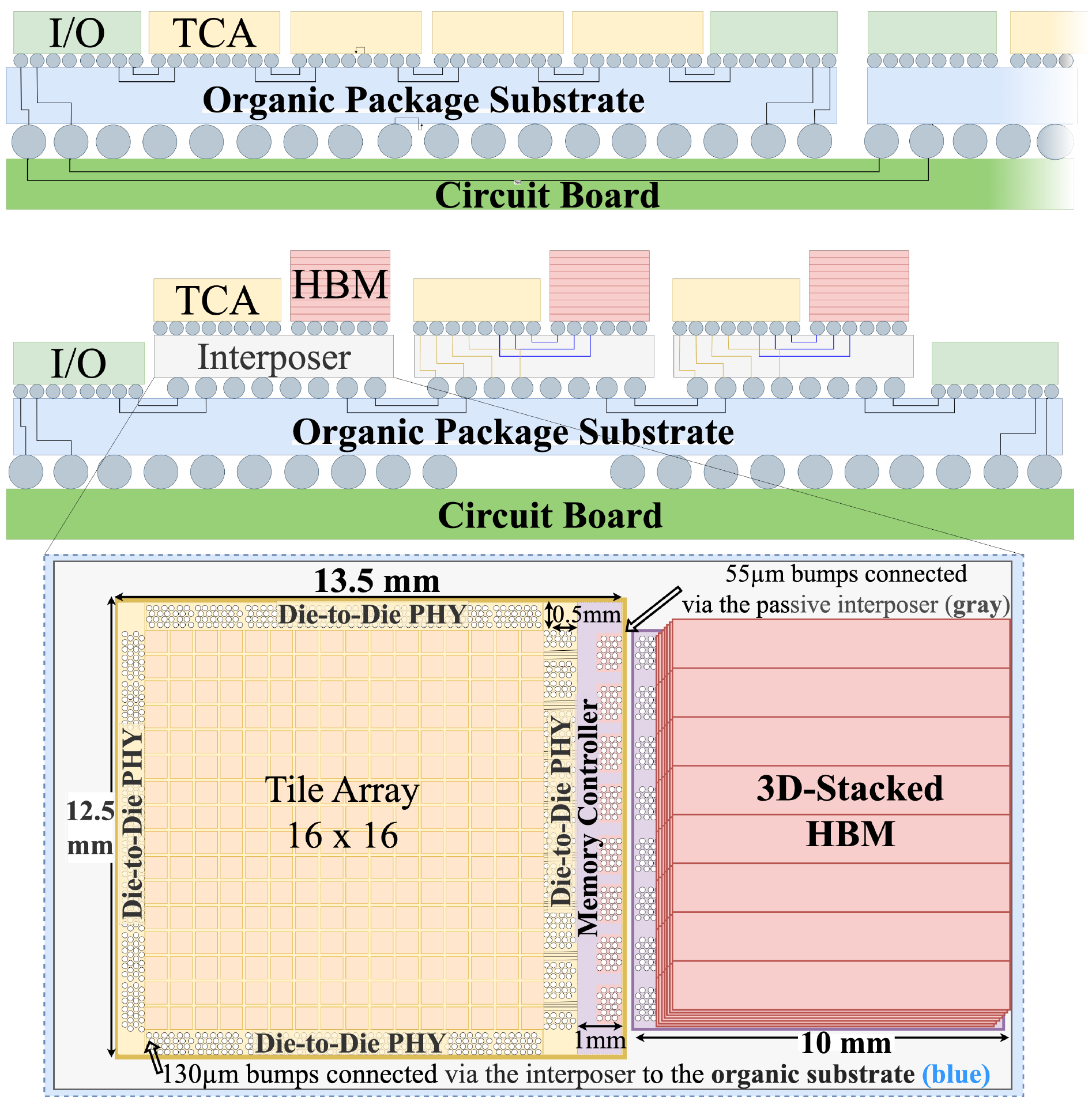}
\vspace{-4mm}
\caption{
Two possible integrations of \proj. Top: two packages on a board, each featuring only tiled-compute array (\device) dies, optimized for time-to-solution as it maximizes parallelization (lower data footprint per die).
Bottom: a single package with \device dies and stacked DRAM, optimized for performance-per-watt/\$ (\cref{sec:results}).
}
\vspace{-2mm}
\label{fig:cake}
\end{figure}

In the last decade, we have seen the rise of massive manycore systems~\cite{cerebras,groq,tesla_dojo,tpu_google,nvidia_a100} that target AI workloads via dataflow computation.
There is, however, still an unmet demand for systems that can massively accelerate applications that utilize sparse data structures~\cite{iarpa_agile}.

Recent works have proposed task-based parallelization schemes that accelerate graph and sparse data traversal by splitting the program at irregular memory accesses~\cite{fifer,polygraph,dalorex, jeffrey_hive}, offering a promising path for parallelizing these communication- and data-intensive applications. 
While some solutions (e.g. Dalorex~\cite{dalorex}) scale up to thousands of processing units (PUs) on-chip, the challenge of parallelizing these applications across millions of cores has not been tackled.

Starting with a tile-based architecture, in order to achieve such scaling, one must address the following challenges:
~(1) \emph{Scalable long-distance communication}:
Due to the irregular nature of the communication in graph applications, the larger the number of processing elements, the higher the average communication distance. This increases both latency and contention on the network.
~(2)~\emph{Work imbalance}:
Because of the power-law distribution of many graphs, growing dataset sizes result in a growing number of edges for a small number of vertices.
When parallelizing across more cores, this, in turn, results either in serialization due to atomics and coherence overheads (e.g. in do-all parallelization schemes) and/or work imbalance (e.g. single-owner-per-data parallelization scheme by Dalorex) when scaling out.
~(3)~\emph{Energy, performance, and cost efficiency trade-offs}:
Building a system with millions of cores requires the design of a multi-node system that brings with it various additional design and performance constraints. In order to fully analyze trade-offs of energy, performance, and cost efficiency, a detailed architectural design of a full multi-node system with complete communication primitives for fine-grained messages across multiple chips is necessary.
 
These challenges have not been addressed so far with previous studies on accelerating sparse and graph applications.
Even for the manycore systems that are currently available, we lack detailed studies of large system-design tradeoffs.

\textbf{Our approach:} We present a multi-node architecture for massively parallel execution of sparse and graph applications.
Our \textul{ta}sk-oriented \textul{s}calable \textul{c}hiplet \textul{a}rchitecture for \textul{d}istributed \textul{e}xecution system (\proj), overcomes the above limitations by:
(a) utilizing an atomic-free data-local execution model that utilizes proxy ownership of data, relaxing the \emph{single-owner-per-data} constraint employed by Dalorex by allowing proxy tiles to merge updates to data owned by distant tiles---thereby reducing work-imbalance and network traffic;
(b) designing a scalable system with nodes composed of grids of \textul{t}iled \textul{c}ompute \textul{a}rray (\device) chiplets, optionally interleaved with DRAM as shown in Fig.\ref{fig:cake};
and (c) studying tradeoffs in cost, performance, and energy of packaging-time design decisions.

We evaluate strong scaling performance when parallelizing sparse applications with up to one million processors across 256 nodes, with no dataset preprocessing or partitioning.

\textbf{\emph{Proxy Regions and Selective Cascading:}}
\proj shards the dataset across the tile grid allocated to run a program, similar to Dalorex.
In addition, to reduce long-distance communication, \proj divides the tile grid into subgrids.
In each subgrid, each tile is assigned proxy ownership to a chunk of data as if the entire data array were distributed only within this subgrid. 
Proxy owners are permitted to cache an update to remotely stored/owned original data.
From here on in this paper, we call the tile that owns the original data array chunk the ``owner tile'' to distinguish it from the proxy owners.
Tasks requiring long-distance communication are first sent to a proxy owner within the sender's region.
If an update needs to be sent to the owner tile eventually, additional proxy owners en route may also process the task as proxy tasks.
This is triggered opportunistically: the system moves the invocations to the owner tile directly when no network traffic is present in order to minimize staleness but utilizes all available proxy owners when they are idle, and there is network traffic.
When applied to tasks that perform associative operations, this results in accumulating updates in a cascading manner and filtering out otiose invocations at the owner tile, significantly reducing the average communication distance of task invocations (see \cref{sec:results}).
This dynamic approach also improves work balance and network contention, as many of the task invocations targeting a tile that owns very hot data are filtered out by the proxy tiles.

\textbf{\emph{Scalable System Design:}}
Designing systems composed of millions of cores is only practical if cost constraints are taken into account.  We investigated the fabrication cost of data-local architectures like Dalorex and found that the granularity of SRAM per tile affects key metrics of throughput and energy efficiency per dollar. 
Using these metrics, we studied pre-silicon and packaging-time choices---not done in previous studies.
Because fabricating separate silicon masks for different target metrics is expensive, we propose a composable architecture where crucial decisions like on-chip memory capacity or off-chip bandwidth are made after silicon fabrication.
Fig.\ref{fig:cake} depicts two packaging options where the same \device{} dies are integrated to create different chip products, each optimized for a different metric.
A \device{} die may be attached to a DRAM device during packaging.
This effectively increases the capacity of a tile's \textit{local memory} since the DRAM storage exclusively assigned to the tiles of the adjacent \device{} die.

\noindent
\textbf{The technical contributions of this paper are}:
\begin{itemize}
\itemsep0em 
    \item Software-reconfigurable proxy ownership and dynamic selective cascading that allow aggregation of updates from distant tiles efficiently, which improves work balance and reduces network contention. 
    \item A scalable architecture where chip design decisions such as on-chip memory, \# of PUs, network topology and off-chip bandwidth are made post-silicon.
    \item A viable path for interleaving any \# of columns of compute and DRAM chiplets on a chip package.
    \item A detailed study of tradeoffs in cost, performance, and energy efficiency of several package designs.
\end{itemize}

\noindent
\textbf{We evaluate \proj and demonstrate that:}
\begin{itemize}
\itemsep0em 
    \item \proj exhibits strong scaling up to 1 million PUs, while prior work starts to plateau after thousands.
    \item The proxy region improve performance by 2.6$\times$\ geomean over Dalorex~\cite{dalorex} for 64x64 (single chip package), and by 3.3$\times$\ geomean when scaling to larger grid.
    \item Optimized task scheduling and buffering improve the overall system's throughput up to 4.7$\times$.
    \item Different packaging options of \proj lead to different optimal design points for key metrics in HPC, such as energy efficiency and throughput per unit cost.
    \item Our BFS parallelization is 3.4$\times$ faster than the most performant entry of the Graph500 list for RMAT-26, and 5.2$\times$ faster for RMAT-22.
\end{itemize}

\section{Background and Motivation}\label{sec:background}

Memory accesses at graph and sparse linear algebra applications do not exhibit spatial or temporal locality, resulting in poor cache behavior and intense traffic in the memory hierarchy~\cite{graphattack}.
Prior work aiming to accelerate these workloads mitigate memory latency via decoupling, prefetching, and hardware pipelining techniques~\cite{prodigy,maple,graphicionado,graphpulse,ozdal,chronos, pipette, fifer, polygraph, jeffrey_hive, jeffrey_swarm}.
Fifer~\cite{fifer} and Polygraph~\cite{polygraph} increase utilization further through spatiotemporal parallelization, while Hive~\cite{jeffrey_hive} provides ordered parallelization.
However, all of these works store the dataset on off-chip DRAM; thus, their scalability is limited by the network and memory bandwidth.
Tesseract~\cite{tesseract} and GraphQ~\cite{2019graphq} tackle this problem via processing-in-memory.
However, their proposed integration of PUs on the logic layer of a memory cube~\cite{hmc} is less scalable than the manycore integration proposed by Dalorex~\cite{dalorex}.


We detail the data-local execution model of Dalorex in \cref{sec:background_dalorex} since this paper 
extends this model to improve work balance and network contention, as well as to design a more practical system integration.
We then review the latest manycore architectures in \cref{sec:background_manycore} with a focus on chip manufacturing.

\setlength{\tabcolsep}{1.5pt}
\begin{table}[t]
\centering
\small
\begin{tabularx}{\columnwidth}{@{\hspace{-1pt}}l@{\hspace{-14pt}}c@{\hspace{1pt}}c@{\hspace{2pt}}c@{\hspace{1pt}}c@{\hspace{1pt}} }
\toprule
 \textbf{Reconfigurable}                                   & Memory               &  \#Processing                       &  Configurable            & Software ISA \\
 \textbf{Aspects} /                                      & Capacity             &  Elements                          &  Network                 & Programmable \\
 Prior Work                                              & on package          &  on package                                &  Topology                & Processors \\
\midrule
Fifer \cite{esperanto}                                & {\color{mustard}\pmb{\cmark}} & {\color{darkred}\pmb{\xmark}} & {\color{darkred}\pmb{\xmark}} & {\color{darkred}\pmb{\xmark}} \\ 
Tesseract~\cite{tesseract},GQ~\cite{2019graphq}      & {\color{darkgreen}\pmb{\cmark}} & {\color{darkred}\pmb{\xmark}} & {\color{darkred}\pmb{\xmark}} & {\color{darkgreen}\pmb{\cmark}} \\
Dalorex \cite{dalorex}                                & {\color{darkred}\pmb{\xmark}} & {\color{darkred}\pmb{\xmark}} & {\color{darkred}\pmb{\xmark}} & {\color{darkgreen}\pmb{\cmark}} \\
PolyGraph \cite{polygraph}                            & {\color{mustard}\pmb{\cmark}} & {\color{darkred}\pmb{\xmark}} & {\color{darkred}\pmb{\xmark}} & {\color{darkred}\pmb{\xmark}} \\
Decades~\cite{piton+ariane,decades}                   & {\color{darkred}\pmb{\xmark}} & {\color{darkred}\pmb{\xmark}} & {\color{darkred}\pmb{\xmark}} & {\color{darkgreen}\pmb{\cmark}} \\
\midrule
ESP~\cite{giri_date20}        & {\color{darkred}\pmb{\xmark}} & {\color{darkred}\pmb{\xmark}} & {\color{darkred}\pmb{\xmark}} & {\color{darkgreen}\pmb{\cmark}} \\
Manticore \cite{manticore}                            & {\color{darkgreen}\pmb{\cmark}} & {\color{darkred}\pmb{\xmark}} & {\color{darkred}\pmb{\xmark}} & {\color{darkgreen}\pmb{\cmark}} \\
Intel's PIUMA \cite{piuma}                            & {\color{darkgreen}\pmb{\cmark}} & {\color{darkred}\pmb{\xmark}} & {\color{darkred}\pmb{\xmark}} & {\color{darkgreen}\pmb{\cmark}} \\
Graphcore \cite{graphcore}                            & {\color{darkred}\pmb{\xmark}} & {\color{darkred}\pmb{\xmark}} & {\color{mustard}\pmb{\cmark}} & {\color{mustard}\pmb{\cmark}} \\
Sambanova \cite{sambanova}                            & {\color{mustard}\pmb{\cmark}} & {\color{mustard}\pmb{\cmark}} & {\color{darkgreen}\pmb{\cmark}} & {\color{darkred}\pmb{\xmark}} \\
Cerebras \cite{cerebras}                              & {\color{darkred}\pmb{\xmark}} & {\color{darkred}\pmb{\xmark}} & {\color{darkgreen}\pmb{\cmark}} & {\color{darkgreen}\pmb{\cmark}} \\
Groq \cite{groq}                                      & {\color{darkred}\pmb{\xmark}} & {\color{darkred}\pmb{\xmark}} & {\color{darkred}\pmb{\xmark}} & {\color{darkgreen}\pmb{\cmark}} \\
Tesla Dojo \cite{groq}                                & {\color{darkgreen}\pmb{\cmark}} & {\color{darkgreen}\pmb{\cmark}} & {\color{darkred}\pmb{\xmark}} & {\color{mustard}\pmb{\cmark}} \\
Esperanto \cite{esperanto}                            & {\color{darkred}\pmb{\xmark}} & {\color{darkred}\pmb{\xmark}} & {\color{darkred}\pmb{\xmark}} & {\color{darkgreen}\pmb{\cmark}} \\ 
Google TPU~\cite{tpu_google}                          & {\color{mustard}\pmb{\cmark}} & {\color{darkred}\pmb{\xmark}} & {\color{darkgreen}\pmb{\cmark}} & {\color{darkred}\pmb{\xmark}} \\ 
Nvidia A100~\cite{nvidia_a100}                        & {\color{darkgreen}\pmb{\cmark}} & {\color{darkred}\pmb{\xmark}} & {\color{darkred}\pmb{\xmark}} & {\color{darkgreen}\pmb{\cmark}} \\ 
\midrule
\proj                                             & {\color{darkgreen}\pmb{\cmark}} & {\color{darkgreen}\pmb{\cmark}} & {\color{darkgreen}\pmb{\cmark}} & {\color{darkgreen}\pmb{\cmark}} \\
\bottomrule
\end{tabularx}
\caption{Post-silicon configurability of the on-chip network topology, memory capacity and \# of processing elements (PEs) per chip package, and whether the PEs can execute software instructions.}
\label{table:related}
\end{table}

\subsection{The Data-Local Execution Model}\label{sec:background_dalorex}

\textbf{\emph{Tasking and queuing:}}
In Dalorex, the original program is split into tasks that are executed at the tile co-located with the memory region that the task operates on.
A task can spawn its dependent tasks by injecting inputs for each task in the output queue (OQ) of the tile it is executed at, which drains into a logical network channel.
At each tile, there is one input queue (IQ) per task type for incoming task invocations. 
An IQ is populated with invocation parameters that are either (a) directly pushed by a prior task executed in the local tile, or (b) coming as task messages from the network channels. 

\textbf{\emph{Task prioritization:}}
In Dalorex, work efficiency and PU utilization are highly impacted by the order of task executions.
The task scheduling unit (TSU) determines the order of execution of tasks based on the occupancy of queues.
It prioritizes tasks whose IQ is highly populated, or OQ is empty.
The TSU ``senses'' the network pressure and executes tasks that will relieve pressure when it is high (IQs full) or increase when it is low (OQs rarely pushing).
\proj uses \emph{reconfigurable queue sizes}, which \cref{sec:queues} demonstrates to be beneficial since different applications perform best with distinct queue sizes.

\textbf{\emph{Distant task communication:}}
A program in data-local execution is a sequence of tasks invoking other tasks upon pointer indirection, with no concept of execution threads (no main task).
Thus, spawned tasks may target any random tile in the grid, i.e., whoever owns the data to be processed next.
As the size of the grid increases, so would the average number of router hops of a task message and, thus, the network contention.
\proj introduces \emph{proxy ownership} (see \cref{sec:proxy_cascading}) which relaxes the constraint of \emph{single-owner-per-data} in Dalorex.
\cref{sec:res_dalorex} shows how proxy improves performance by reducing the excessive traffic occurring in large tile grids.

\textbf{\emph{Tile grid and network routing:}}
The size of the grid a program runs on is determined by the user at compile time.
Thus, the logical tile ID is microcoded when a workload is selected to run on a given grid, and it is used for XY routing.
Since the dataset is statically partitioned across the tiles on the grid, and the first parameter of a task message is a global index to a data array, this index is used to route the message, avoiding message headers altogether.
Based on the size of the array associated with a network channel for routing purposes, the router selects the bits that indicate the destination tile ID.
\proj supports \emph{selective cascading} by allowing proxy tiles to grab task messages despite not being the destination tile.

\subsection{Fabrication Cost of Manycore Systems}\label{sec:background_manycore}

In recent years, the widespread demand for large deep-learning models has accelerated the development of manycore and dataflow systems that can massively parallelize compute-intensive workloads.
Although the demand for graph and sparse linear algebra workloads is growing, we have not seen systems exhibiting as high strong-scaling performance as dense systems due to their irregular data-access patterns and low arithmetic intensity (ops/byte).

Table~\ref{table:related} shows a selection of the recent manycore systems where the upper ones focus on sparse workloads and the rest focus on AI.
While some of these manycores have good attributes for sparse data processing (large on-chip SRAM capacity~\cite{dalorex,cerebras_fft,groq,graphcore,tesla_dojo} or on-package DRAM~\cite{nvidia_a100,sambanova,tpu_google,manticore,tesseract}), the ratio of memory capacity and PUs is optimized for dense computation, and the interconnect is designed for dataflow communication.
For our work, instead of designing a manycore architecture with a memory-to-PU ratio and network optimized solely for sparse data, we propose a chiplet architecture where these can be configured at chip packaging time; we evaluate them of in \cref{sec:packaging} also considering cost-effectiveness.

\textit{\textbf{Packaging-time configurability for better silicon reuse:}}
The semiconductor industry is increasingly utilizing having multiple dies in a chip package~\cite{ibm_telum,shapphire_rapids,ponte_vecchio,amd_rome,amd_vcache} to reduce fabrication costs and enable reusing components across different products.
From the silicon manufacturing perspective, the Non-Recurring-Engineering (NRE) cost of a wafer mask is so high with respect to the silicon wafer itself that the cost per wafer is 18$\times$ larger when manufacturing 100 wafers, instead of 100,000~\cite{lithovision}.
Since the volume of silicon production of each chip architecture generation is not as high in the HPC market as in the consumer electronic market, we argue that the mass fabrication of a chiplet that serves as a building block for scalable architectures is a valuable proposition.
Thus, we designed a chiplet that can be fabricated at high volume to amortize NRE costs, and then integrated into a Multi-Chip Module (MCM) package as an arbitrary grid of dies.
Fig.\ref{fig:cake} shows that \proj allows integrating DRAM between columns of \device{} dies (and not just on the chip edges as in prior work).

\section{The Tascade Approach}\label{sec:approach}


\cref{sec:proxy_cascading} describes how \proj overcomes the scalability challenges of prior work through proxy ownership and selective cascading.
\proj also offers several \emph{software-configurable} features that can be tuned for each application and task to optimize performance.
Particularly, \cref{sec:sw_reconfig} presents these features for task scheduling, queue sizes and Network-on-Chip (NoC), while
\cref{sec:cache_config} describes our approach for caches and prefetching, and
Table~\ref{table:configuration_knobs} summarizes the remaining configurable knobs.
\cref{sec:package_reconfig} introduces our \emph{package-time configurable} design and provides a thorough discussion of some packaging options.

\subsection{Proxy Owners and Selective Cascading}\label{sec:proxy_cascading}
Programs that can be parallelized for loop iterations have the underlying assumption that the order of iterations, as well as interleavings of operations across different iterations, preserves correctness.
Many graph and sparse applications have commutative operations, making them amenable to such parallelization, provided that all writes to the same data are atomic.
This is the underlying power of models such as Bulk-Synchronous Parallelization (BSP) to guarantee eventual correctness allowing interleavings of all other read/write operations.
Distributed-memory MapReduce~\cite{mapreduce} implementations may avoid atomic operations by updating copies of the result array and merging them at the end.
However, since pre-merge computations and merging cannot be effectively overlapped in such software schemes, its leads to idle PUs.
This under-utilization is exacerbated in the context of BSP, where each epoch has a barrier.
The data-local programming model proposed by Dalorex eliminates the need for atomic updates with a \emph{single-owner-per-data} task-based model. However, large parallelizations using this model suffer from work imbalance as only a single tile's PU can operate on a given data, and work-per-data is highly skewed.

Tascade relaxes the single-owner-per-data constraint of Dalorex and employs \textit{\textbf{two modes of data ownership}}.
(1)~\emph{Original data owner}:
As in Dalorex, dataset arrays are sharded as equal-sized chunks to tiles, where each tile \emph{owns} a chunk of each data array;
(2)~\emph{Proxy data owner}: 
We allocate \emph{proxy regions} by subdividing the tile grid into smaller regions.
For selected tasks, we distribute the proxy ownership of copies of the data array(s) across the tiles of \emph{each} proxy region, such that the most recent updates coming from within the region can be stored in the tile's proxy cache (P\$, detailed in \cref{sec:cache_config}).
Proxy regions allow intermediary storage to perform reduction operations of remote data through \emph{proxy tasks}.
So, in addition to commutativity, \proj leverages the \emph{associativity} in BSP operators: data updates are merged in any combination with eventually correct convergence. 

\begin{figure}[t]
\centering  
\includegraphics[width=\columnwidth]
{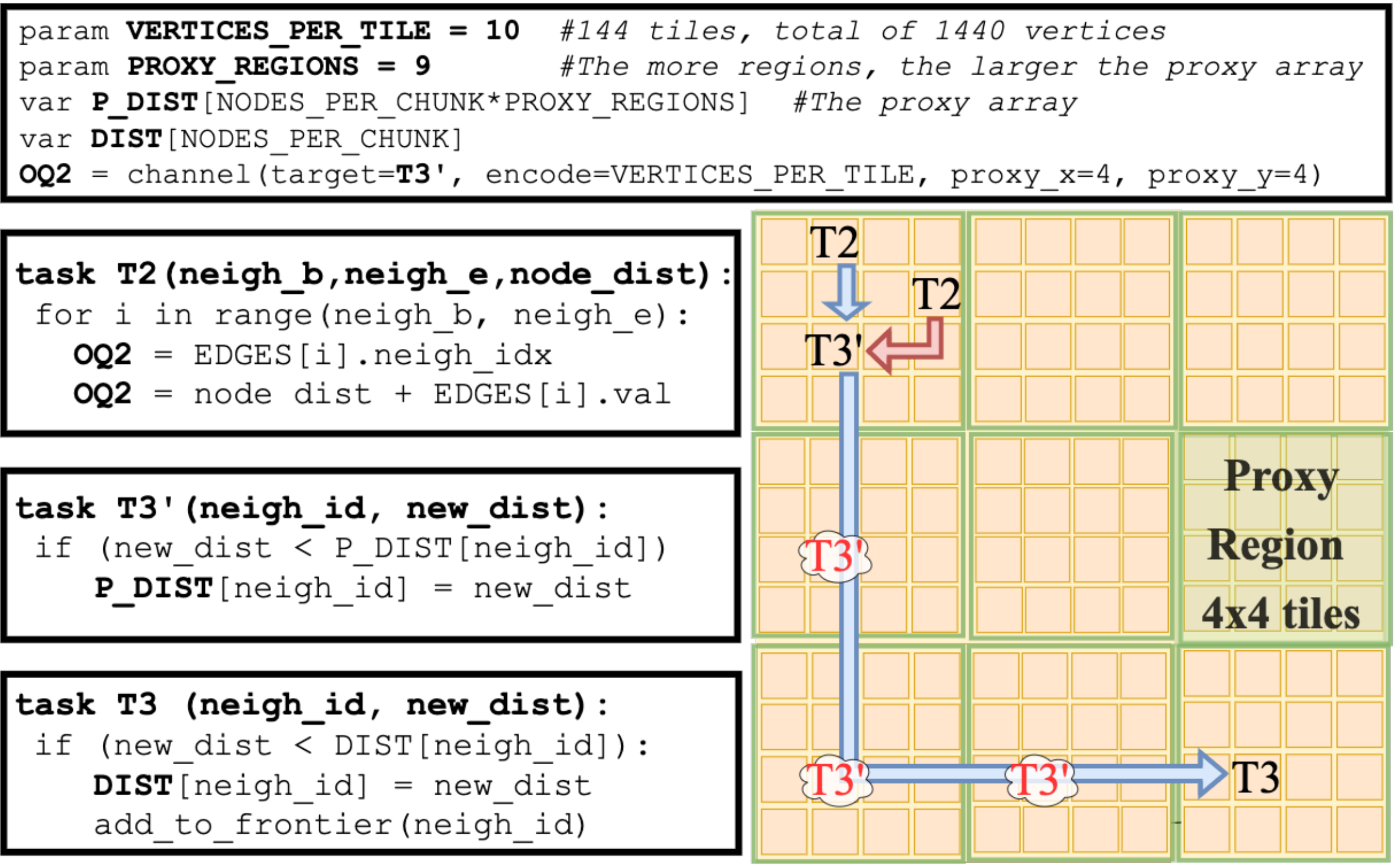}
\caption{
Software configuration of proxy regions and flow of tasks for single-source shortest path (SSSP).
This grid example configured 9 proxy regions of 4x4 tiles.
Each region contains proxy owners for an entire data array (P\_DIST), where each tile \emph{is a proxy for} a fraction of it.
Contrary to Dalorex, in Tascade, T2 invokes T3', which targets a proxy tile within the same region.
T3' tasks write to P\_DIST, which lives in the proxy cache (P\$, see \cref{sec:cache_config}).
The \textcolor{blue}{blue arrow} shows a T3' that results in an eviction in P\$ and triggers a T3 invocation to the data owner.
The \textcolor{red}{red arrow} shows a T3' that does not result in an update in P\$.
Through selective cascading, en route to the owner, this message can also invoke T3' at a local proxy indicated by clouds when there is NoC contention.
}
\label{fig:proxy}
\end{figure}

\textbf{\textit{Applicability:}}
Fig.\ref{fig:proxy} explains how proxy regions are configured in software.
Each proxy region is responsible for a copy of a dataset array.
Any task in the data-local execution model can have a proxy task.
In our evaluation (\cref{sec:results}), we apply this to the vertex update task of graph applications and the reduction phase of the histogram and sparse algebra workloads.
Proxy tasks can be applied to read-only data as well as to modifiable arrays.
For the latter, aggregated data updates are eventually sent to the owner, resulting in eventual consistency.
Fig.\ref{fig:proxy} depicts this scenario with two proxy tasks (T3') being invoked on the same tile, coming from different T2 tasks.

\textbf{\textit{Hardware support for reduction operations:}}
Data updates to a proxy tile eventually trigger a task invocation towards the owner tile.
This happens transparently to software, thanks to the write-back P\$.
When a P\$ cacheline is evicted, the data is 'written-back' by invoking another proxy task in the direction of the owner tile.
For that, the P\$, similarly to the PU, has the ability to push into the OQs.
This hardware support enables \proj to keep the perks of Dalorex's single-owner-per-data task model while improving work-balance by allowing multiple tiles to perform updates.
Moreover, since \proj enables overlapping updates to proxy P\$s with merges with the original data array, it avoids the synchronization of software-based implementations that use data array copies.

\textbf{\textit{Cascading updates:}}
Since proxy ownership within each region is allocated in the same way, the proxy tiles for a particular array element are on the same row/column for horizontally/vertically aligned proxy regions.
Therefore, as a task invocation travels towards the owner tile across the NoC, it will naturally pass by its  corresponding proxy tiles en route (Fig.\ref{fig:proxy}, blue arrows).
The router of a proxy tile has the option to grab a proxy task if the tile is free and there is NoC contention on the port towards the owner tile.
This alleviates NoC traffic and increases load-sharing.
We call this \emph{selective cascading}.
Proxy tiles closer to the owner tile are more likely to have the most updated value, leading to the filtering of unnecessary updates in a cascading manner.

\textbf{Takeaway 1.} Proxy regions and selective cascading provide the following advantages:
(a)~\textit{smooth out work imbalance} by allowing multiple tiles to operate on a given data; and
(b)~\textit{minimize the \# of bytes traversing the NoC} by coalescing updates to the original data array en route;
(c)~\textit{reduce NoC contention} by opportunistically deciding whether proxy tasks are executed at proxy tiles, or they continue towards the owner.

\subsection{Software-configurable Queues and NoC}\label{sec:sw_reconfig}
\textbf{\textit{Task Scheduling and Queue Sizes:}}\label{sec:taskQueueSizes}
To mitigate back-pressure from end-point routers, input queues must have a sufficient size to buffer message bursts.
These bursts naturally happen when a column of a sparse matrix is much denser than the rest.
Conversely, making queues too large can cause messages to wait too long on average to be processed, leading to data staleness.
Since there is no global orchestration of tasks, scheduling solely depends on information that is local at each tile.
Thus, queue occupancy and back-pressure help signal tiles to adapt their task scheduling to prioritize tasks leading to higher overall utilization, effectively resulting in a Goldilocks effect.
\proj queue sizes are configurable in software.
We study the performance impact of queue sizes in \cref{sec:queues}.
%
We observe that optimal queue sizes are application-specific, which is related to applications' sensitivity to data staleness. 
We envision that workload- and task-specific queue sizes can be set by a compiler via static analysis and the communication distance between tasks.
This can also help utilize the storage better and give more SRAM to the caches (\cref{sec:cache_config}).


\textbf{\textit{NoC-to-task Mapping:}}
In a \proj system, there are $P$ physical NoCs and $C$ logical channels. 
These channels can be assigned to different physical NoCs or the same one.
If they are in separate NoCs, these messages are routed uncontested.
If they are in the same NoC, they are arbitrated in a round-robin fashion.
The allocation of physical to logical resources is fully reconfigurable, a NoC can be used by multiple tasks, and a task can use multiple NoCs.
If an application is not using all of the available NoCs, a task can use more than one channel to increase the bandwidth for this task.


\textbf{Takeaway 2.} Software-configurable queues enable selecting a sufficient size that helps maintain a software pipeline effect in this task-based throughput-oriented architecture while minimizing the SRAM resources utilized for that.
The software mapping of task channels to physical NoCs enables optimally allocating network resources given the consumption and production rate of certain task types.

\begin{scriptsize}
\setlength{\tabcolsep}{1pt}
\begin{table}[t]
\centering
\small
\begin{tabularx}{\columnwidth}{l}
\toprule
\textbf{Tapeout-time Design Decisions} \\
\midrule
\# of Tiles per die \\
SRAM per-tile (MiB) \\
\#Physical NoCs and Width of each one (in bytes)\\
Max.\# of Router Buffer Entries per Physical NoC \\
Max.\# of Logical NoC Channels per Physical NoC \\
Max.\# of Memory Controller Channels per \device{} die \\

\midrule
\textbf{Packaging-time Design Decisions} \\
\midrule
\# of \device{} dies per package\\
\# of DRAM dies per package and capacity of each (GiB)\\
\device{}-DRAM ratio dictates Mem. BW and Capacity per \device{} die\\
\# of I/O dies per package (Off-Package BW) \\
\midrule
\textbf{Compile Time Configurations} \\
\midrule
Size and Place of the grid that the workload uses (grid of dies)\\
Whether the dataset uses SRAM as a scratchpad or D\$ \\
Size of the D\$ (in data elements) \\
Mapping of Tasks to Logical NoC Channels  \\
Mapping of NoC Channels to Physical NoCs \\
\# of Router Buffer Entries of each NoC Channel (shared pool)\\
Arbitration ratio between Channels sharing a Physical NoC\\ 
\midrule
\textit{Per Task} \\
\midrule
If a Task has a Proxy and the size of the Proxy region \\
If the Proxy array is cached or kept in full \\
If a Task uses data prefetching during execution (if cached dataset)\\
Input Queue (IQ) Size \\
Output Queue (OQ) Size if the Task produces into a NoC Channel \\
\bottomrule
\end{tabularx}
\caption{Reconfigurable Parameters of \proj}
\label{table:configuration_knobs}
\end{table}
\end{scriptsize}

\subsection{Software-configurable Caches}\label{sec:cache_config}

\proj designs the SRAM in each tile such that it can be used either as a scratchpad or as a cache~\cite{amoeba,cameo,esp_cache}.
The cache mode stores cacheline tags in SRAM, including a valid bit.
To minimize the hardware and energy overhead of using the SRAM as a cache, we make the caches directly mapped.

This cache mode is used to configure the \emph{proxy cache} (P\$), which holds proxy data, and the \emph{data cache} (D\$), which holds original (exclusive owner) data arrays. 
When scaling out the parallelization of a dataset, if the memory footprint per tile is low-enough to fit in the local SRAM, the D\$ would not be configured, the PUs would access the data arrays as a scratchpad, and the memory controller and the HBM would be switched off.
The footprint of a P\$ decreases when the size of the proxy region increases (the performance impact of region sizes is studied in Fig.~\ref{fig:proxy_comp}).
The D\$ and the P\$ have reconfigurable sizes.
When multiple tasks use proxies, each task would configure its own logical P\$, although they all use the same comparison logic.
Only one D\$ can be configured, and its line width equals the bitline width of the DRAM memory controller (512 bits in our experiments with HBM).

\textbf{\textit{D\$ misses and evictions:}}
Upon a miss, the D\$ fetches the full cacheline from DRAM without checking for coherence since the data is not shared.
Each tile's local SRAM is backed up by a DRAM storage of size $DRAM\_capacity / tiles\_per\_die$.
(The details of the DRAM devices assumed for the evaluation are described in Table~\ref{table:wire_param}.)
The D\$ has one dirty bit per line to write back to DRAM upon eviction.
Since the D\$ of each tile only contains the part of the dataset that the tile is responsible for, there are no coherence issues for modified data.

\textbf{\textit{P\$ misses and evictions:}}
A miss in the P\$ returns a default predefined value such as 0 for Histogram or infinite for SSSP.
On eviction, the data is simply replaced in a write-back manner, and so the evicted data is sent as a task targeted to the data owner.
In our experiments, a P\$ line contains one element to avoid sending multiple updates upon eviction.

\textbf{\textit{Prefetching:}}\label{sec:prefetching}
The PU has a very simple in-order pipeline, which stalls waiting for data on a D\$ miss.
Since the first parameter of every task message contains an array index, and the TSU knows to which array it corresponds (used for NoC routing), we use this information to prefetch the data.
A task may access more than one array using the same index, so we add another pointer to the TSU's per-task table so that it prefetches both when needed.
Those tables also have an extra bit saying whether the PU keeps prefetching data during the task execution.
Since pointer indirections are split into tasks, when tasks access multiple array elements, they often do it with a streaming pattern, 
To prefetch these, we enable PU's next-line prefetcher for tasks that access multiple elements.

\textbf{Takeaway 3.}
Software-configurable caches enable utilizing the SRAM efficiently by balancing the resources dedicated to D\$ and P\$, depending on what is their footprint. E.g., when sharding a dataset across more tiles, its footprint per tile decreases, and more SRAM can be dedicated to P\$. 

\begin{figure}[t]
\centering
\includegraphics[width=\columnwidth]{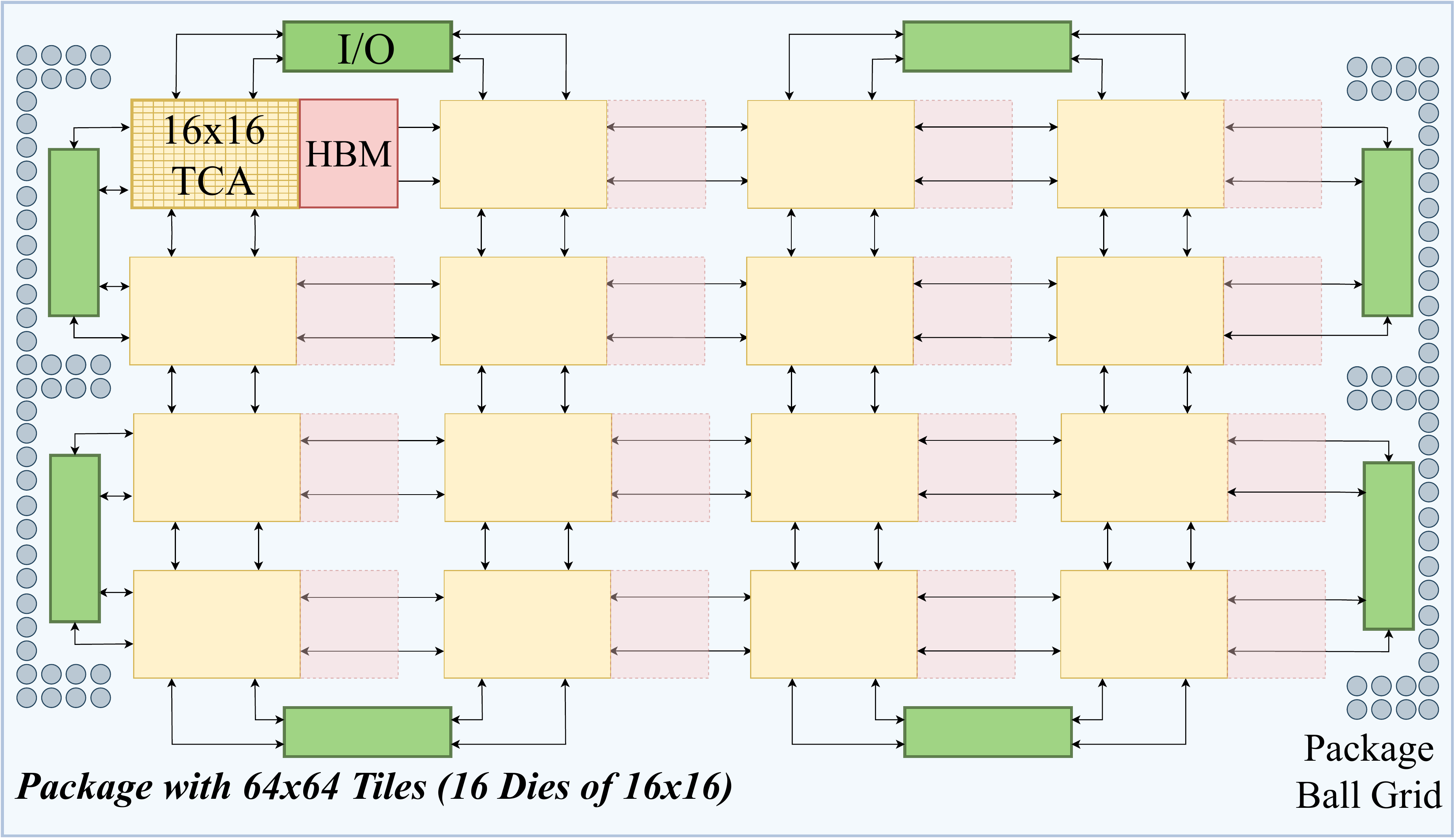}
\caption{
Top view of an example chip package with 64x64 tiles.
The number of \device{} dies would determine the compute capacity, while including HBM dies would determine the compute-to-memory ratio of the chip architecture.
The number of I/O dies (and their resources) would determine the provisioned off-chip bandwidth.
}
\label{fig:package}
\end{figure}

\begin{figure*}[t]
\centering  
\includegraphics[width=\textwidth]{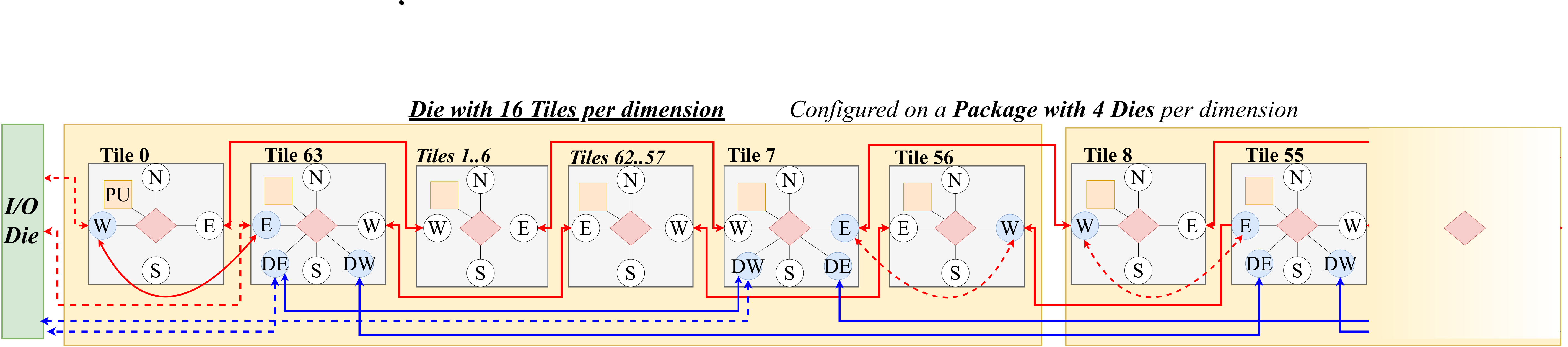}
\caption{
Horizontal links within a \device{} die and across dies.
The \textcolor{red}{red links} show the NoC that connects every tile (\textcolor{red}{\textit{tile-NoC}}), while the \textcolor{blue}{blue links} show the NoC that connects to one tile per die (\textcolor{blue}{\textit{die-NoC}}).
Because of the die-NoC, the routers at the die edges are radix-9, while the rest are radix-5.
The ports shadowed in blue are runtime reconfigurable; any tile subgrid within the system may become torus (including across packages).
The dies on the edges of a package will interface with the \textcolor{darkgreen}{I/O die}.
All I/O links are configured when loading the dataset to maximize I/O bandwidth.
During program execution, both NoCs may become torus, or tile-NoC torus, and die-NoC mesh, to keep streaming from I/O.
}
\label{fig:torus}
\end{figure*}

\subsection{Packaging-time configurability}\label{sec:package_reconfig}

\textit{\textbf{Memory technology and integration:}}
The memory hierarchy determines the bandwidth available to computing units and, thus, at which level of arithmetic intensity the chip becomes memory-bound.
To increase memory bandwidth, prior work has proposed having DRAM on the chip.
Some proposals assumed 3D integrations where processing cores are on the base plane of the 3D-stacked memory, accessing the TSVs directly~\cite{tesseract,2018graphp,2019graphq,pim_hbm}.
However, this may render impractical.
The are very few foundries that can fabricate HBM technology, and the design seems to only change with every generation of JEDEC specification~\cite{jedec_2015} (e.g., HBM2, HBM2e, HBM3) which vendors conform to.
So it would be expensive to fabricate an HBM die with a custom base layer.
Moreover, the base layer of the HBM stacks is filled with MBIST logic and PHY~\cite{hbm2_sk_hynix,hbm3_sk_hynix,hbm,hbm_samsung}.
Alternatively, other works have proposed integrations of off-the-shelf HBM dies with compute dies by having the compute dies surrounded by HBM devices on the edges of the chip~\cite{amd_epyc_isca,amd_rome,emib,kaby_lake,shapphire_rapids,manticore,nvidia_chiplets,nvidia_a100}.
\proj is the first proposal for having horizontally-integrated HBM dies interleaved across columns of compute dies. 
This allows the design of \device{} dies to be agnostic to the \# of \device{} and DRAM dies that are eventually integrated on-chip.

\textbf{\textit{Interleaving DRAM \& \device{} dies:}}
Figs.\ref{fig:cake} and \ref{fig:package} depict the integration of \device{}, and HBM dies via a passive silicon interposer.
This provides higher \device{}-HBM bandwidth than a silicon bridge~\cite{emib} or substrate integration~\cite{flipchip_options,amd_rome}.
The interposer only contains the wires between a \device{} die, and its HBM die since HBM access is exclusive. The links connecting \device{} dies are routed through the organic substrate, which also contains the power delivery and redistribution layer (RDL).

\textbf{\textit{How much DRAM \textul{can} be included?}}
HBM capacity can be increased by either stacking more DRAM layers or having larger layers.
In our evaluation, we assumed a 10x11mm 4-layer HBM die (up to 16 layers have been demonstrated~\cite{hbm3_sk_hynix}).
One could also integrate larger HBM dies, provided that the distance between \device{} dies remain within 25mm (limit of common PHYs for MCM~\cite{ucie,bow,die2die_comp}).

\textbf{\textit{How much DRAM \textul{should} be included?}}
In addition to the arithmetic intensity of the target application, selecting the amount of DRAM on the chip depends on the expected level of parallelism for the final product.
If the chip is always expected to parallelize a dataset to the limit of strong scaling (having little workload per tile), HBM might not be necessary.
In contrast, with target metrics like performance-per-watt, larger on-package memory capacity is essential.
We evaluate both options in \cref{sec:packaging} and discuss the tradeoffs.


\textbf{\textit{Memory controller:}}
As depicted in Fig.~\ref{fig:cake}, the memory controller lives on the \device{} die, and thus, in case of not integrating DRAM, it becomes \textit{dark silicon}. Although in our evaluation, we consider it as such, an idea for a future design would be to have an embedded FPGA logic that allows hardening a memory controller or other computing logic (in the case DRAM is not integrated). Another advantage is enabling a DRAM technology like DDR (which may use a different memory controller than HBM) to tradeoff capacity for bandwidth.

\textbf{\textit{Designing the NoC:}}
When dealing with irregular traffic in dimension-ordered routing (DOR), a 2D torus provides a more uniform utilization than a 2D mesh~\cite{dalorex}.
DOR keeps the router logic simple, and deadlocks are avoided using bubble routing~\cite{bubble}.
To make implementation practical on 2D silicon, the 2D torus is folded: one set of links connects the even-numbered tiles, and the other set of links connects the odd-numbered ones.
The torus can be arbitrarily large, i.e., be confined within a die or span multiple dies, packages, or boards.
Table~\ref{table:wire_param} presents the interconnect energy and latency assumed for our evaluation at each level.
Fig.\ref{fig:torus} shows that the routers at the edges of each die can be configured to connect to a router on the next die or wrap around by connecting to the adjacent tile (Tiles 0 and 63).

\textbf{\textit{Reconfigurable topology:}}
We can reconfigure a 2D-torus NoC into two 2D-mesh NoCs by not connecting the wrap-around links on the routers at the edges.
Utilizing two mesh NoCs may be the right choice when running a workload with near-neighbor communication~\cite{paralle_computing_dwarfs}.
As Fig.\ref{fig:torus} shows, \proj has two hierarchical NoCs, one that connects every tile and one that hops once per die.
Each NoC topology is individually configured.
While bringing the data inside the package from disk, they both would be configured as a mesh to enable data streaming from the I/O dies.
During the execution, both may become torus, or the die-NoC can remain open to I/O data.

\textbf{\textit{Off-Chip Bandwidth:}}
Our design uses I/O chiplets to connect chip packages. This allows \device{} dies to remain agnostic to a specific off-chip protocol (e.g., PCIe 6.0~\cite{pcie6}).
It also defers choices like off-chip bandwidth to packaging time, so they are taken based on the requirements of the final product.
The off-chip bandwidth could be as high as the I/O-\device{} bandwidth.

\textbf{\textit{What decides the package size?}}
A package is as big as the size of its substrate.
A silicon substrate typically scales 2-3x the reticle size, but this could technically scale up to a wafer-scale interposer~\cite{kumar_wafer}.
On a Multi-chip Module (MCM) integration, server-class chip packages typically span over $4500mm^2$~\cite{amd_epyc_isca,amd_rome,shapphire_rapids}.
However, MCM substrates have been fabricated to nearly $100,000mm^2$~\cite{tesla_dojo,big_chip_comparison}.
So the choice of the package size may be determined by the unit of sale rather than a constraint in the substrate size.
Regarding the number of dies on a package, although we see chip packages with tens of them~\cite{ponte_vecchio}, bonding may set feasibility bound.
The bonding yield of each die determines the overall package yield. 
~\cite{cost_model_package}.
When using \proj dies of 16x16 tiles, the practical limit may be 128x128 tiles per package (64 dies).
In our evaluation, we use a chip containing 64x64 tiles (shown in Fig.~\ref{fig:package}); it spans $4780 mm^2$, which is in line with the size of server-class chips.

\textbf{\textit{Scaling beyond a package:}}
Off-chip interconnect topology and bandwidth are determined by the I/O chiplets selected.
The I/O links across packages could also be electrical or optical~\cite{mcm_optical,photonic_chiplets}.
If a system featuring \device chiplets is expected to have near-neighbor communication, several packages can simply be connected on a board using a 2D interconnect with electrical links.
However, when irregular or all-to-all traffic is expected, it would be beneficial to connect packages with optical links on a low-diameter interconnect~\cite{benner2012optical,blue_gene,slimf}.
Despite our evaluated workloads being irregular, for simplicity, we assumed a 2D network with electrical links across packages.

\textbf{\textit{Data partitioning}} is a preprocessing step used in distributed graph processing to minimize cross-node communication~\cite{giraph++,metis}.
In addition to proxies, \proj could potentially use partitioning and place the data so that each subgrid holds a partition. However, our evaluation did not need to resort to it.

\textbf{Takeaway 4.}
\proj proposes a chiplet-based architecture that allows making key design decisions, such as on-chip memory and off-chip interconnect, post-silicon.
This enables the same \device{} design to be mass-produced (saving NRE costs) and later integrated differently to create products with varying target metrics.
We have discussed cases that benefit from one configuration or another---some of which are evaluated in \cref{sec:results}.

\section{Evaluation Methodology}\label{sec:methodology}

\textbf{\textit{Applications:}}
We evaluate the performance of \proj on four graph workloads, one sparse linear algebra, and a histogram benchmark~\cite{parboil} to demonstrate the generality of our approach for memory-intensive applications.
\textit{Breadth-First Search (BFS)} determines the number of hops from a root vertex to all vertices reachable from it;
\textit{Single-Source Shortest Path (SSSP)} finds the shortest path from the root to each reachable vertex;
\textit{PageRank} ranks websites based on the potential flow of users to each page~\cite{pagerank};
\textit{Weakly Connected Components (WCC)} finds and labels each set of vertices reachable from one to all others in at least one direction (using graph coloring~\cite{connected_components});
\textit{Sparse Matrix-Vector Multiplication (SPMV)} multiplies a sparse matrix with a dense vector.
\textit{Histogram} counts the \# of values that fall within a series of intervals.

\textbf{\textit{Datasets:}}
We use various sizes of the RMAT~\cite{kron} graphs---standard on the Graph500 list~\cite{graph500}---e.g., 
RMAT-26 (abbreviated as R26 in \cref{sec:results}) contains $2^{26}$ vertices (V)
and 1.3B edges (E), and has a memory footprint of 12GB.
We also use the Wikipedia (WK) graph (V=4.2M, E=101M) in our evaluation.

\textbf{\textit{Simulation:}}
Our evaluation uses muchiSim~\cite{muchisim}, the cycle-level manycore simulator that was used to evaluate prior work Dalorex~\cite{dalorex}.
We extended muchiSim to support chiplets and on-chip DRAM; Table~\ref{table:wire_param} summarizes our additions to the energy, latency, and area model for communication links and memory technology.
We assume the same logic frequency as Dalorex, 1GHz.
We also added a cost model to study the cost-effectiveness of different \proj configurations.

\textbf{\textit{Cost Model:}}
For \textit{silicon}, we assume that a 300mm wafer with a 7nm transistor process costs \$6,047~\cite{lithovision}.
We obtain the cost per die by dividing the wafer cost by the number of good dies per wafer (calculated using Murphy's model~\cite{yield_calc} with 0.2mm scribes, 4mm edge loss, and 0.07 defects per $mm^2$).
When comparing cost-effectiveness in Fig.\ref{fig:packages_comparison}, we do not include the Non-Recurring Engineering (NRE) cost of the \device{} dies since all the options use the same \device{} chiplets.
In terms of \textit{packaging}, all our results featuring grid sizes over 64x64 use multiple packages (of 64x64 tiles each, based on our rationale from \cref{sec:package_reconfig}).
We assume the cost of the 65nm silicon interposer connecting a \device{} die with HBM (including bonding) to be 20\% of the price of a \device{} die~\cite{cost_model_package}; the cost of an organic substrate to be 10\% of the price of an equal-sized \device{} die, and the bonding to add an additional 5\% overhead~\cite{cost_model_interposer,bonding_yield}.
Regarding \textit{DRAM}, we assume an 8GB HBM2E device with eight 64GB/s memory channels.
While this cost is not disclosed, we made an educated guess using public sources~\cite{lithovision,cost_hbm}.
We assume 7.5\$/GB, which is more affordable than when HBM was released in 2017. One could expect this price to decrease over time as more vendors fabricate HBM~\cite{micron_hbm,hbm2_sk_hynix,hbm_samsung}.

\begin{table}[t]
\centering
\small
\begin{tabularx}{\columnwidth}{@{\hspace{0pt}}l@{\hspace{-25pt}}r@{\hspace{0pt}}}
\toprule
\textbf{Memory Model Parameters} & \textbf{Values} \\
\midrule
SRAM Density\ & 3.5 MiB/mm$^2$~\cite{renesas_ff7nm} \\
SRAM R/W Latency \& E. & 0.82ns \& 0.18 / 0.28 pJ/bit~\cite{renesas_ff7nm} \\
Cache Tag Read \& cmp. E. & 6.3 pJ~\cite{renesas_ff7nm,ariane_cost}\\
HBM2E 4-high Density\ & 8GiB/110mm$^2$ (75 MiB/mm$^2$)~\cite{hbm2_sk_hynix} \\
Mem.Channels \& Bandwidth & 8 x 64 GB/s~\cite{hbm2_sk_hynix} \\
Mem.Ctrl-to-HBM RW Latency \& E. & 50ns \& 3.7pJ/bit~\cite{fine_grain_dram,hbm_samsung_power} \\
Bitline Refresh Period \& E. & 32ms \& 0.22pJ/bit~\cite{refresh_time_hbm,dram_activation_energy} \\
\midrule
\textbf{Wire \& Link Model Parameters}  & \textbf{Values} \\
\midrule
MCM PHY Areal Density                   & 690 Gbits/mm$^2$~\cite{die2die_comp} \\
MCM PHY Beachfront Density              & 880 Gbits/mm~\cite{die2die_comp} \\
Si. Interposer PHY Areal Density        & 1070 Gbits/mm$^2$~\cite{die2die_comp} \\
Si. Interposer PHY Beachfront Density   & 1780 Gbits/mm~\cite{die2die_comp} \\
Die-to-Die Link Latency \& E.          & 4ns \& 0.55pJ/bit ($<$25mm)~\cite{bow} \\
NoC Wire Latency \& E.              & 50 ps/mm \& 0.15pJ/bit/mm~\cite{pim_hbm} \\
NoC Router Latency \& E.            & 500ps \& 0.1pJ/bit \\
I/O Die RX-TX Latency                   & 20ns~\cite{pcie6} \\
Off-Package Link E.                 & 1.17pJ/bit (80mm)~\cite{nvidia_chiplets}\\
\bottomrule
\end{tabularx}
\caption{Energy (E), bandwidth, latency, and area of links and memory devices assumed for the evaluation.}
\label{table:wire_param}
\end{table}

\textbf{Graph500 methodology:}
To understand where our system stands on the Graph500 list~\cite{graph500}, we follow their guidelines.
They measure separately reading and preparing the graph from the search itself.
We do not perform any dataset pre-processing and directly read the CSR structure from the disk.
Since our goal is to evaluate the performance of our architecture, we focus on the graph search time.
We begin counting cycles when the search key is loaded onto the system and stop counting when the last vertex is visited.
We perform one search and report traversed edges per second as $TEPS=m/time$ where $m$ is the number of edges connected to the vertices in the graph traversal starting from the search key.
Since we also evaluate other workloads than graphs, when we report TEPS for SPMV and Histogram, we consider the number of dataset elements.




\section{Results}\label{sec:results}

This section starts by characterizing \proj's settings, which are---as shown in Table~\ref{table:configuration_knobs}---configurable at various levels.
Particularly, \cref{sec:pre_silicon} is about pre-silicon choices: it shows the \textbf{performance impact} of four \textbf{NoC} options and discusses tradeoffs for \textbf{SRAM capacity} per tile and \textbf{die size}.
\cref{sec:queues} characterizes the impact of the software-configurable \textbf{queue} sizes in application runtime and PU utilization.
\cref{sec:res_dalorex} presents the performance gained from using \textbf{proxy} regions over prior work Dalorex and analyzes different region sizes.
\cref{sec:packaging} studies the \textbf{cost-effectiveness} of \textbf{integrating HBM dies} in terms of performance and energy efficiency per dollar.
Finally, \cref{sec:million} studies strong scaling (by parallelizing R26 for grid sizes ranging from 256 tiles to 1,048,576 tiles) and presents performance-per-watt and per-dollar results.
\cref{sec:million} also compares our BFS results with the most performant entries of the Graph500 list~\cite{graph500} and other works~\cite{simula_graphcore_bfs} for two datasets sizes.

\subsection{Pre-silicon Choices}\label{sec:pre_silicon}

\textbf{\textit{SRAM size:}}
We simulated with SRAM sizes ranging from 128KiB to 4MiB and found 1.5MiB per tile to suffice to hold the program code, task queues, and a performing size of D\$ and P\$.
The P\$ saves intra- and inter-die traffic on the task for which it is enabled, reducing the task's average number of router hops.
The ideal size of the P\$ depends on the ratio of the size of the grid (in which a program is running) and the proxy grid.
The D\$ saves traffic to the HBM device, reducing memory-controller contention and energy.
An appropriate size of D\$ yields a hit rate that is high enough not to saturate the memory channels when the processing units are fully utilized.
Cache sizes are compile-time configurable, as mentioned in \cref{sec:sw_reconfig}.
At 1.5MiB, the tile area dedicated to SRAM is 7$\times$ larger than the area of the router, PU and TSU together.
Since \proj is data-centric, this size is a good tradeoff.

\textbf{\textit{Die size:}}
Our evaluation uses chiplets of 16x16 tiles.
This die size is practical because:
(a)~it matches common sizes of HBM chiplets~\cite{hbm2_sk_hynix,hbm3_sk_hynix,hbm_samsung};
(b)~it allows for larger configurations per package, as the packaging yield depends on the number of dies;
(c)~the lower yield of a 32x32-tile die (27x25mm) results in 62\% less good dies per wafer~\cite{yield_calc}.
Although prior work used large dies with high yield at the expense of adding tile redundancy~\cite{cerebras_hotchips,graphcore}, we find it appealing to have a unit die that is small enough to add flexibility at packaging time.

\begin{figure}[htp]
\includegraphics[width=\columnwidth]{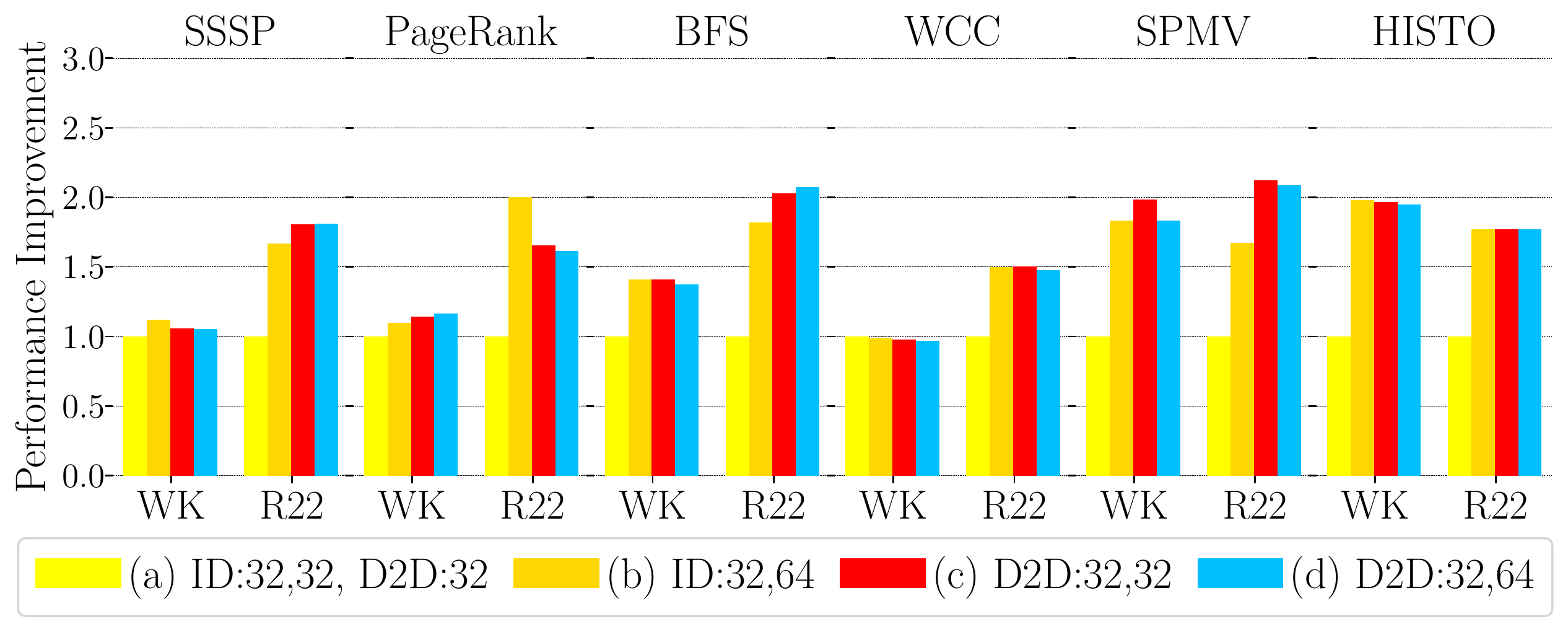}
\caption{Performance improvement of increasing link widths over a baseline of two 32-bit NoCs intra-die (ID) throttled to a shared 32-bit die-to-die (D2D) link.
}
\label{fig:NoC_characterization}
\end{figure}

\textbf{\textit{Studying NoC width:}}
Fig.\ref{fig:torus} showed that \proj uses a die-NoC that hops between dies and a tile-NoC that hops logically adjacent tiles.
We use a single 32-bit die-NoC for all our evaluations. We characterize  how the number and width of the intra-die (ID) and die-to-die (D2D) links of the tile-NoC affect runtime in Fig.\ref{fig:NoC_characterization}, including:
\textit{(a)} two 32-bit links (i.e., two NoCs) whose flow gets throttled to a shared 32-bit link when crossing die boundaries;
\textit{(b)} a 32-bit and a 64-bit link, also throttled to a shared 32-bit link when crossing die boundaries;
\textit{(c)} same as \textit{b} but with two 32-bit links across dies;
\textit{(d)} 32-bit and 64-bit links, both inside and between dies.
We map the channels that deliver the most frequent task type and its proxy (e.g., T3 and T3' in Fig.\ref{fig:proxy}) into the 64-bit NoC (if any),  while other tasks use the 32-bit NoC.

\textbf{\textit{NoC tradeoff:}}
Fig.\ref{fig:NoC_characterization} shows that across the board, the geomean improvement of option \textit{c} with two 32-bit links across dies is 51\% geomean faster than option \textit{a}.
Since the NoC width is determined pre-silicon, it is essential to understand its tradeoffs.
The die area of \textit{c} grows by 4.5\% over that of \textit{a}, but the performance improvement renders it worth it.
\footnote{
The interconnect between \device{} dies in \textit{c} is right below the limit of what is supported with MCM ($\sim$800Gbit/s/mm) to avoid a more expensive interposer needed for (d) that can reach up to 5Tbit/s/mm~\cite{nvidia_chiplets,bow,ucie}.
For the die-to-die substrate wires, we assume BoW-128 PHYs with 130µm bumps~\cite{bow,die2die_comp}.
To connect \device{} and HBM, we assume a 65nm CMOS passive interposer~\cite{65nm_interposer_gf,65nm_interposer_xilinx} with 55µm bumps~\cite{bow}.
}

\textbf{\textit{Choices for the rest of the evaluation:}}
All other experiments in \cref{sec:results} use option \textit{c}.
Note that the characterizations shown in Figs.\ref{fig:NoC_characterization},\ref{fig:queue_characterization},\ref{fig:proxy_comp} evaluate using a 64$\times$64 tile grid that fits within a single chip package of 16 dies (as in Fig.\ref{fig:package}).
All but Fig.\ref{fig:proxy_comp} results use proxy regions of 16x16 tiles, matching the die size.
Thus, there are 16 regions overall in the 64$\times$64 grid.

\subsection{Queue Size Effects on Performance}\label{sec:queues}

\begin{figure}[t]
\includegraphics[width=\columnwidth]{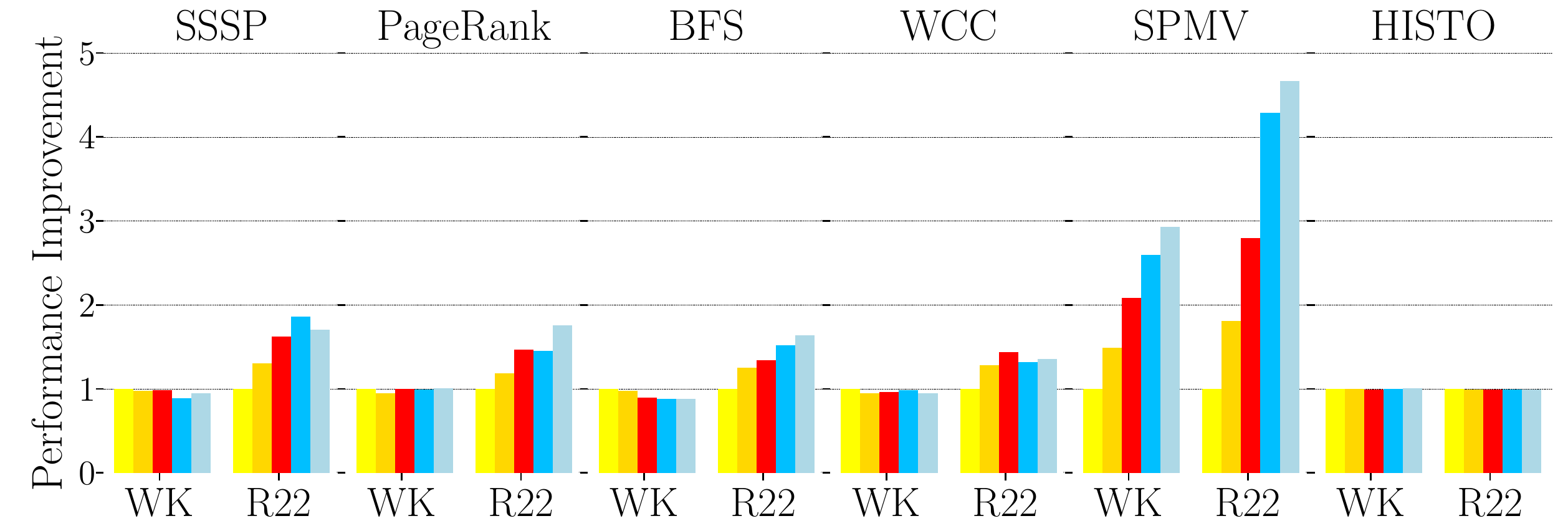}

\vspace{+3mm} \includegraphics[width=\columnwidth]{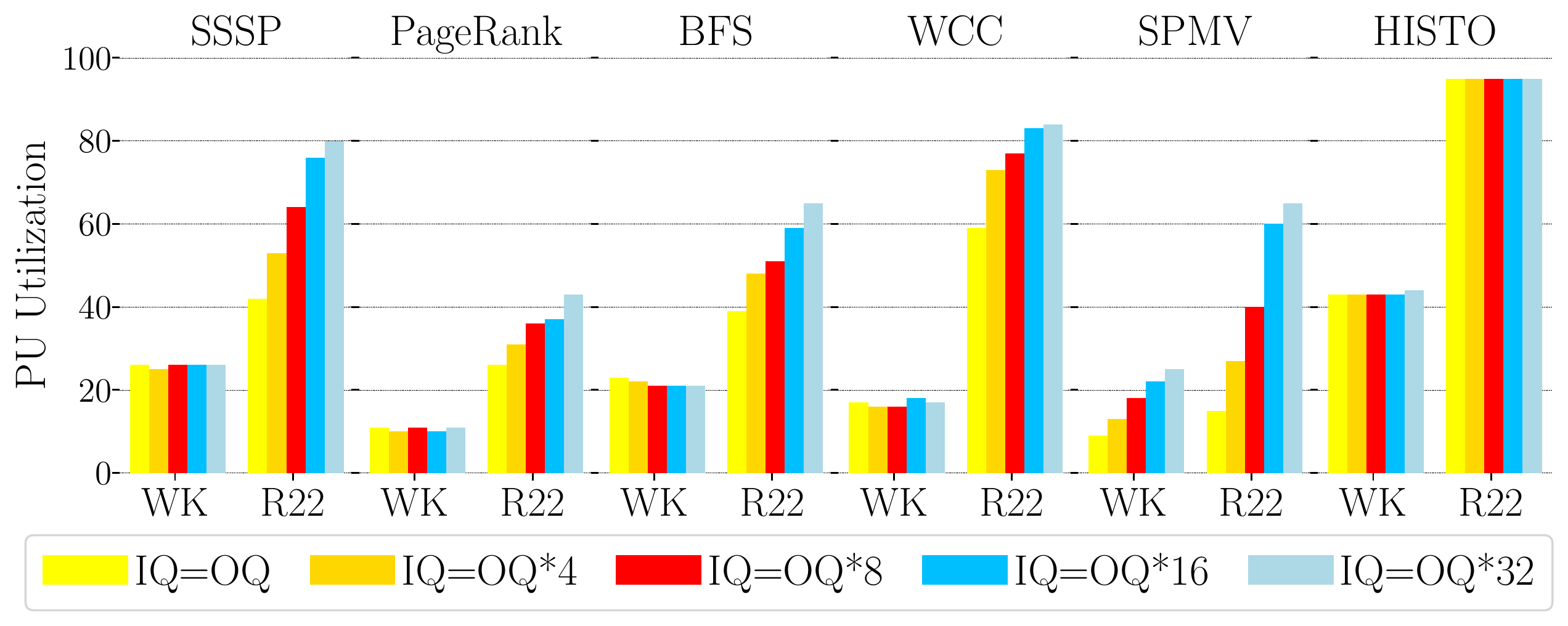}
\caption{Increase in performance (top) and \% PU utilization (bottom) when growing the input queue (IQ) size---keeping the same output queue (OQ)---normalized to a baseline of equal IQ and OQ size.
}
\label{fig:queue_characterization}
\end{figure}

Fig.\ref{fig:queue_characterization} (top) shows the performance gained from increasing the size of IQs (while OQ size is kept constant).
There are tradeoffs associated with queue sizes.
Small IQ sizes lead to end-point contention as message bursts arrive, and the PU is unable to process them quickly enough.
Large IQ sizes contribute to increased data staleness and work redundancy in graph applications and also consume SRAM resources that could be dedicated to caches.
%
Fig.\ref{fig:queue_characterization} shows a peak in the performance of SSSP, BFS, and WCC for smaller queue sizes---as they are most sensitive to staleness---whereas the performance for SPMV improves as the IQ size increases.
The absence of general improvement for WCC and Histogram on K22 can be attributed to the already-high core utilization we observe for these workloads.

Regarding the utilization shown in Fig.\ref{fig:queue_characterization} (bottom), note that the footprint per tile is only $2^{10}$ vertices for R22 executed on $2^{12}$ tiles and only half as many for Wikipedia dataset.
Achieving 100\% utilization is challenging at this extreme level of parallelization because the cool-down phase (during which there is a long tail of a few PUs executing the final tasks) represents a significant portion of the total runtime.

All other experiments in \cref{sec:results} employ an IQ-OQ ratio of 16.

\subsection{Proxy Regions Improve Performance and Scalability}\label{sec:res_dalorex}

\begin{figure}[t]
\includegraphics[width=\columnwidth]{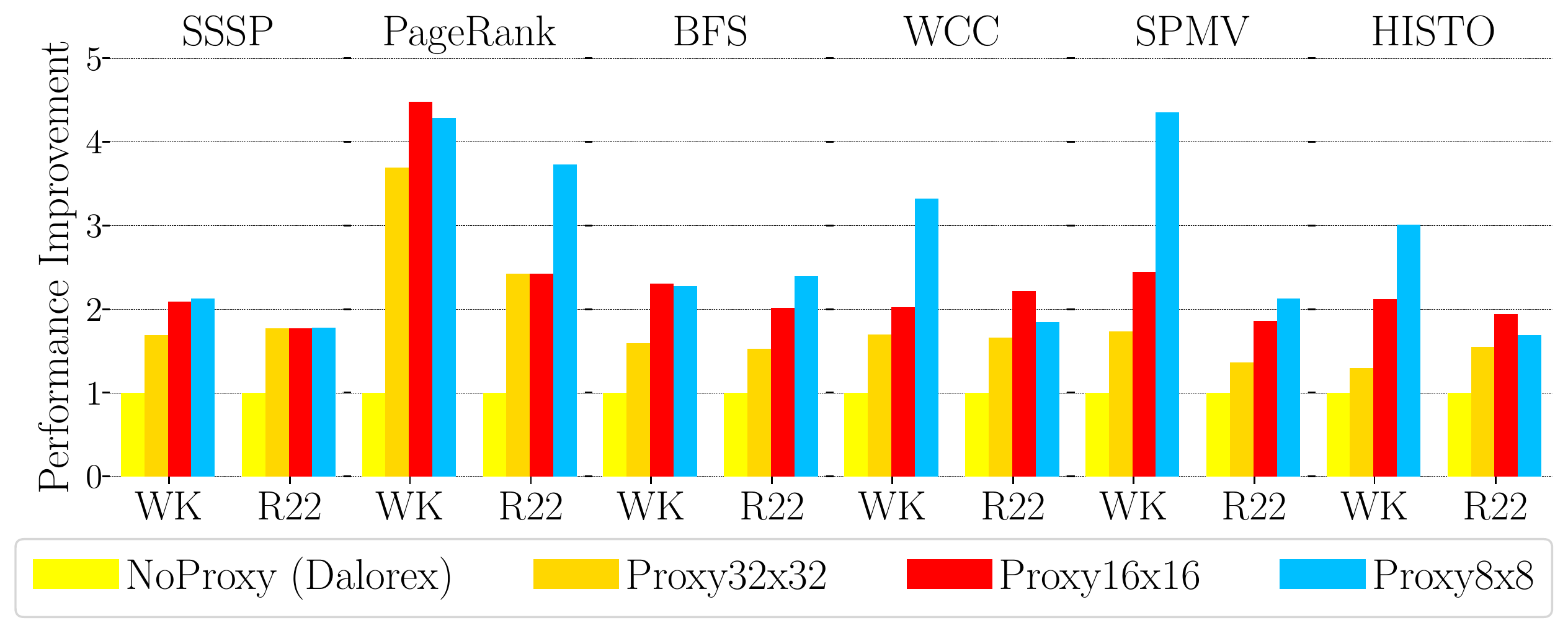}
\caption{
Performance gain with decreasing sizes of proxy regions normalized to the baseline of no proxy, 
}
\label{fig:proxy_comp}
\end{figure}

Fig.\ref{fig:proxy_comp} displays the performance impact of various sizes of proxy regions over a baseline of not using proxies.
This no-proxy setup serves as our method for evaluating Dalorex, as we aim to measure the performance improvements stemming from the proxy regions without considering specific physical implementation design choices.

\textbf{\textit{Proxy advantages:}}
\proj is geomean $2.6\times$ faster than Dalorex (up to $4.4\times$) across datasets and apps.
As \cref{sec:packaging} will show, using proxy regions also improves energy efficiency since updates get coalesced at proxy tiles, and fewer bytes travel all the way towards the data owner.

\begin{figure}[t]
\includegraphics[width=\columnwidth]{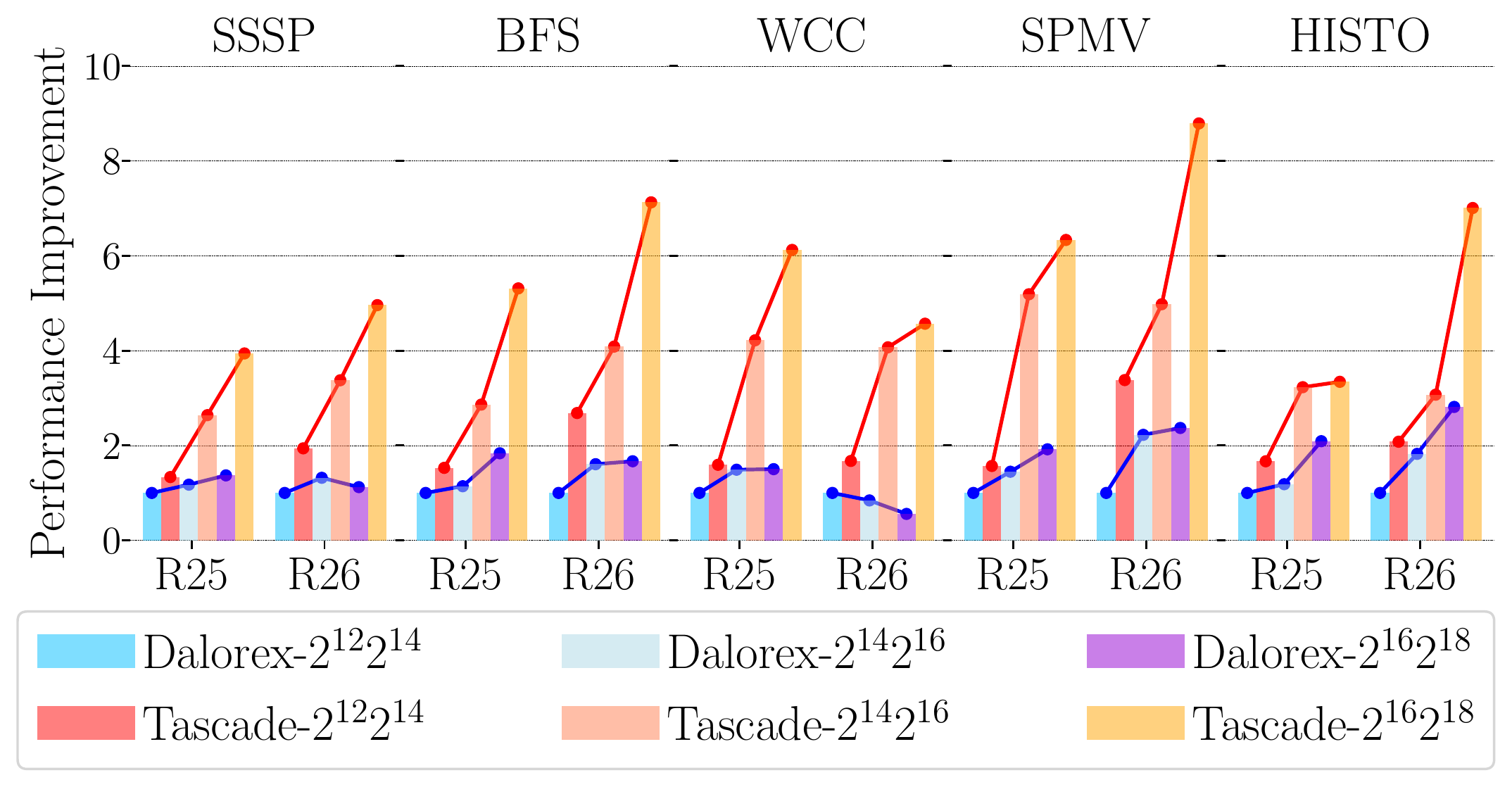}
\caption{
Performance gain of three scaling steps for each dataset.
R25 uses a 64x64 ($2^{12}$ tiles), 128x128 ($2^{14}$) and 256x256 ($2^{16}$) grid for these strong scaling steps, while R26 uses 128x128 ($2^{14}$), 256x256 ($2^{16}$) and 512x512 ($2^{18}$).
}
\label{fig:dalorex_comparison}
\end{figure}

\textbf{\textit{Scalability:}}
Fig.\ref{fig:dalorex_comparison} presents our evaluation of \proj---using proxies of 16x16---and Dalorex on three strong-scaling steps for each dataset, ranging from 64x64 to 256x256 for R25 and from 128x128 to 512x512 for R26.
Note that the starting point is not arbitrary; it is the minimum grid size in Dalorex that can hold the entire dataset on SRAM.
In contrast, \proj can integrate HBM during packaging, and thus, it can run large datasets starting from smaller grids, as we evaluate on \cref{sec:million}.
However, Fig.\ref{fig:dalorex_comparison} compares Dalorex and \proj using the same grid sizes and no HBM.
Dalorex starts to plateau after 64x64 due to high NoC contention, while \proj continues to scale.
For the last scaling step, \proj is on geomean 3.3$\times$ faster than Dalorex across datasets and applications. \proj achieves a speedup of 5.8$\times$ over the first scaling step (and up to 8.8$\times$ for SPMV on R26), while Dalorex only achieves 1.75$\times$ speedup.

\subsection{Package Integration Alternatives}\label{sec:packaging}

\begin{figure}[t]
\subf{\includegraphics[width=\columnwidth]{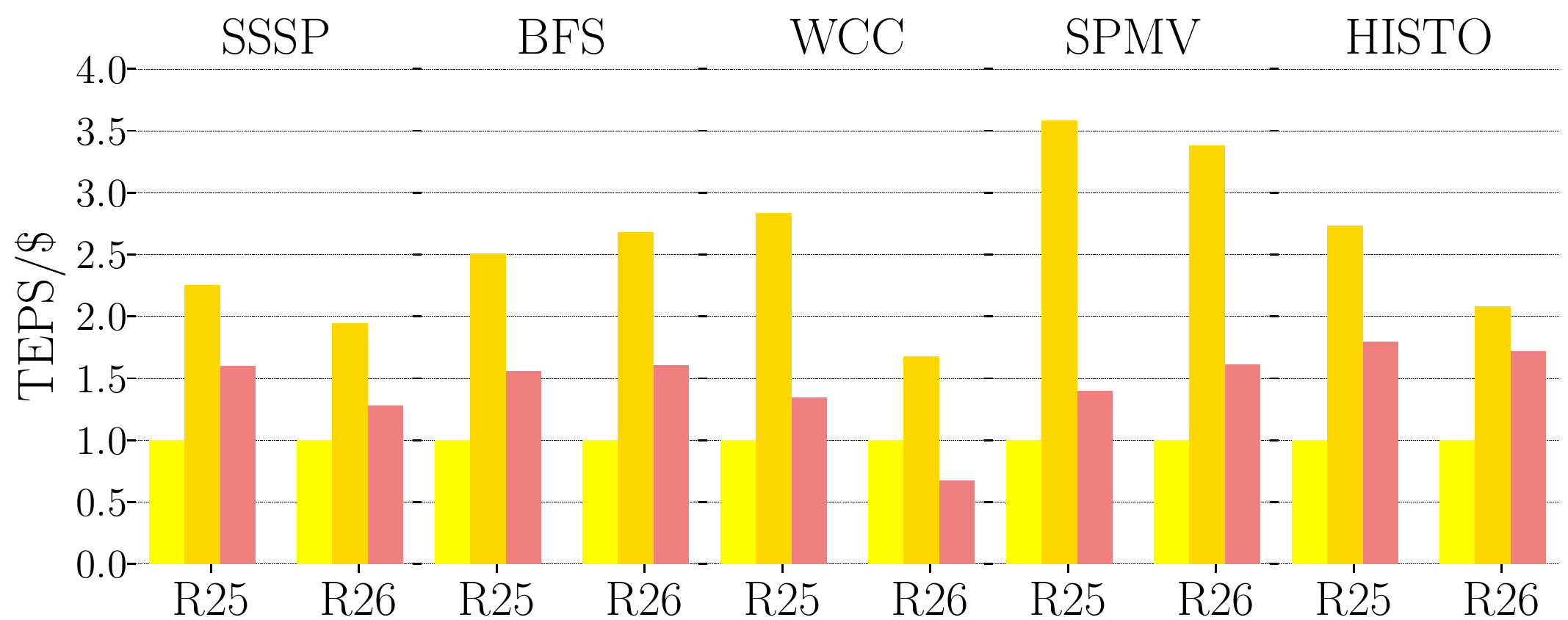} }
\\
\subf{\includegraphics[width=\columnwidth]{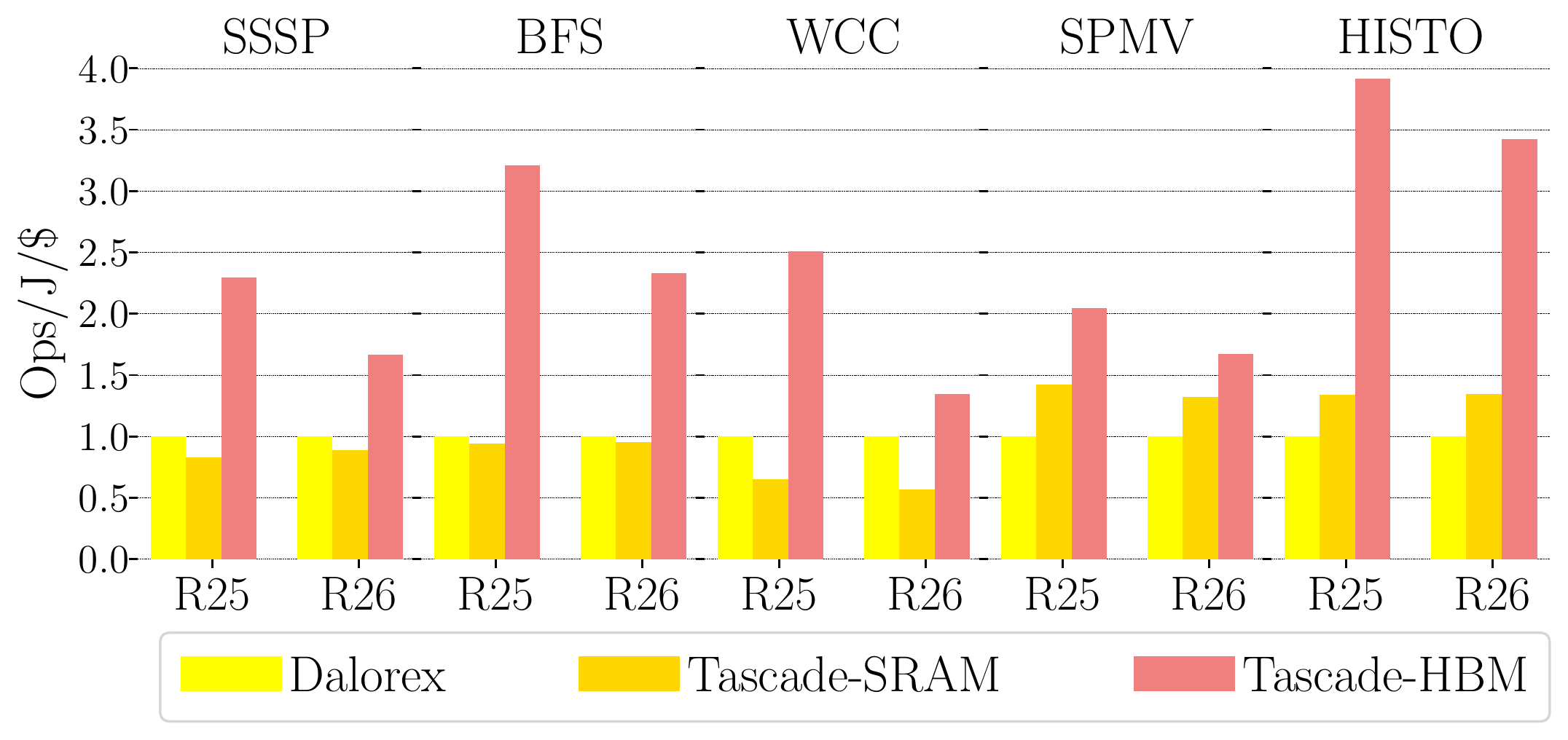} }
{}
\caption{
Normalized improvements in throughput/\$ and energy efficiency/\$.
Dalorex and Tascade-SRAM run on 64x64 tiles (16 dies of 16x16) for R25 and 128x128 for R26, while Tascade-HBM uses 16x16 tiles for R25, and 32x32 for R26.
}
\label{fig:packages_comparison}
\end{figure}

We evaluated throughput and energy efficiency per unit of cost of two \proj integrations over Dalorex (Fig.\ref{fig:packages_comparison}).
As advanced in Fig.\ref{fig:cake}, the \proj integrations we study are (a) having only \device{} dies (and so all memory is on SRAM), and (b) having HBM devices interleaved horizontally between \device{} dies, via a passive interposer.
Since the silicon area of a 64x64 configuration of Dalorex~\cite{dalorex} is beyond the reticle limit of 880mm$^2$, we assume the same chiplet-based integration for Dalorex, as we have done in the previous section.
This makes their cost very similar (only affected by the interconnect area).
Thus, throughput-per-cost differences mostly come from the runtime.
We evaluate all integrations on their smallest configuration that fits the dataset.
This makes the HBM integration use a tile grid of 16 times fewer tiles.

Although Fig.\ref{fig:packages_comparison} does not show absolute throughput, we can observe from Fig.\ref{fig:scaling_million} (also using R26) that it scales.
Since the SRAM configuration of \proj does not cost 16$\times$ more than the one including HBM, \proj-SRAM wins across the board on throughput-per-dollar. However, \proj-HBM wins on energy-efficiency-per-dollar.
This throughput-per-dollar results could change significantly if either HBM becomes cheaper (currently 3$\times$ more expensive than the \device{} die) or another DRAM or 3D-stacked SRAM~\cite{3d_sram} technology becomes cheaper.

\begin{figure}[t]
\includegraphics[width=\columnwidth]{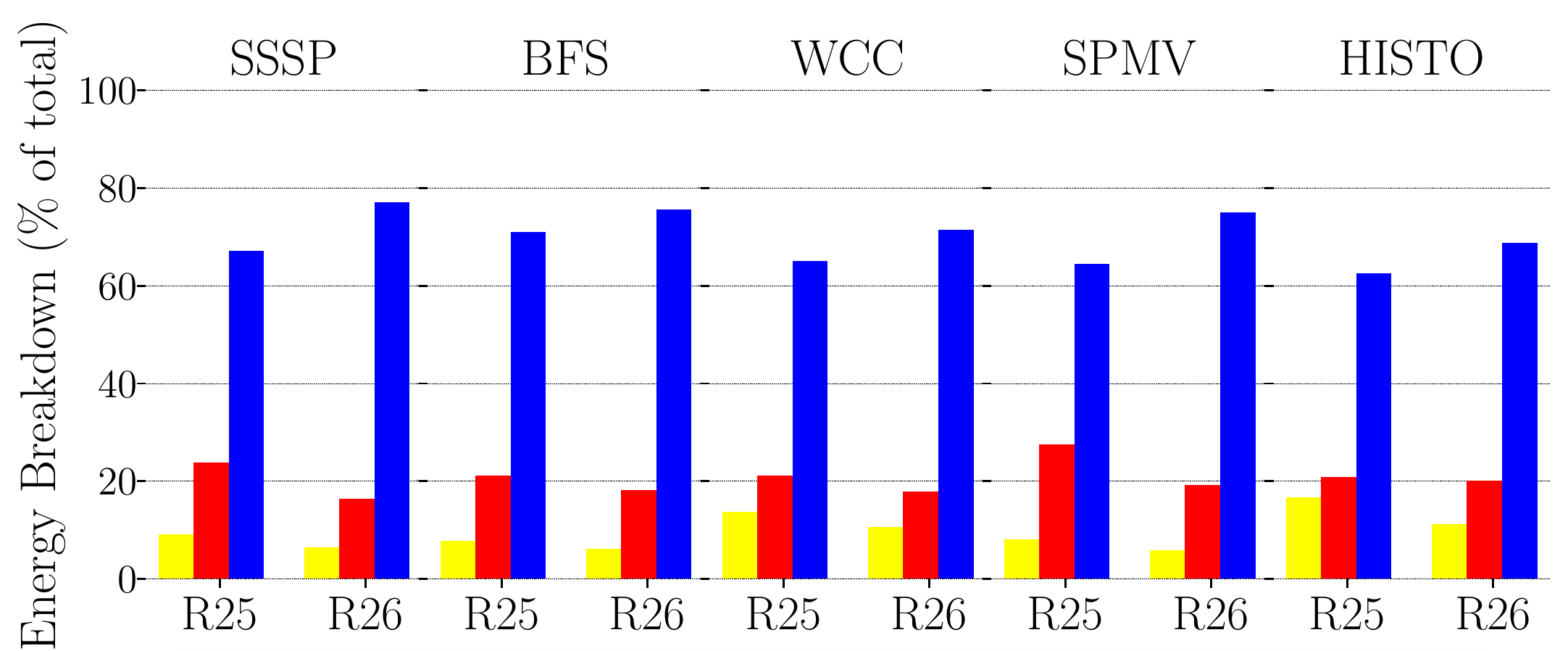}
\includegraphics[width=\columnwidth]{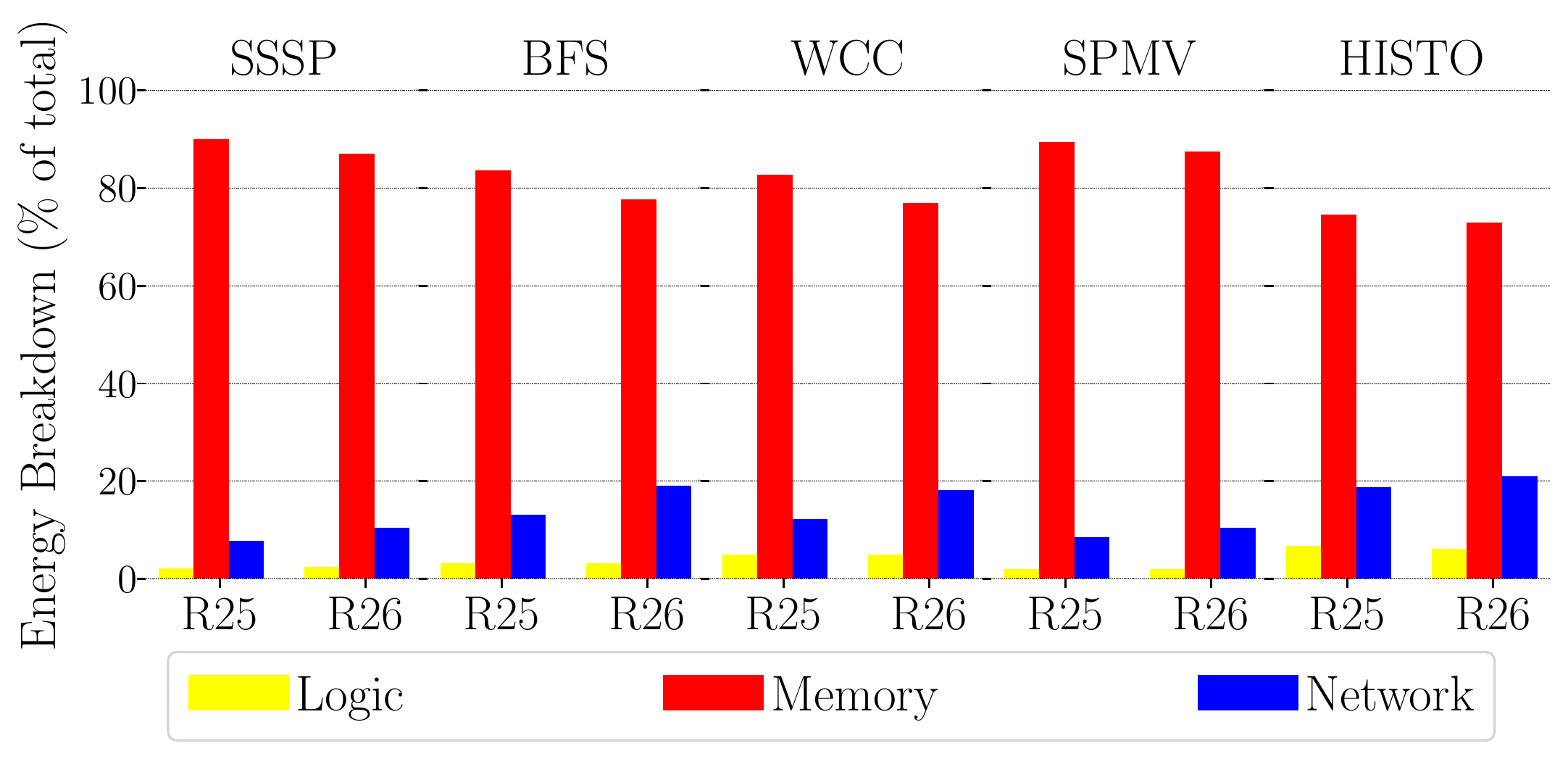}
\caption{
Breakdown of the energy consumed by computing logic, memory, and NoC communication (including routing and wire energy), for Fig.\ref{fig:packages_comparison}'s Tascade-SRAM (top) and Tascade-HBM (bottom).
The Y-axis shows the \% of the total energy spent on each component.
}
\vspace{+1mm}
\label{fig:energy_breakdown}
\end{figure}

We show the breaks down of the energy used by PUs, memory, and NoC for Tascade-SRAM and Tascade-HBM in Fig.\ref{fig:energy_breakdown}.
Since the SRAM-only integration scales out to use 16$\times$ more tiles than Tascade-HBM, it spends more energy on wires.
Although not shown here, we observed that the HBM integration saturates the memory controller bandwidth.
This makes the energy usage dominated by DRAM; PUs use a small fraction in both cases.
Note that PUs are powered off when idle, so they only consume energy while processing tasks.

\subsection{Strong Scaling Up to a Million Tiles}\label{sec:million}
\begin{figure}[t]
\hspace{-3mm} \includegraphics[width=1.05\columnwidth]{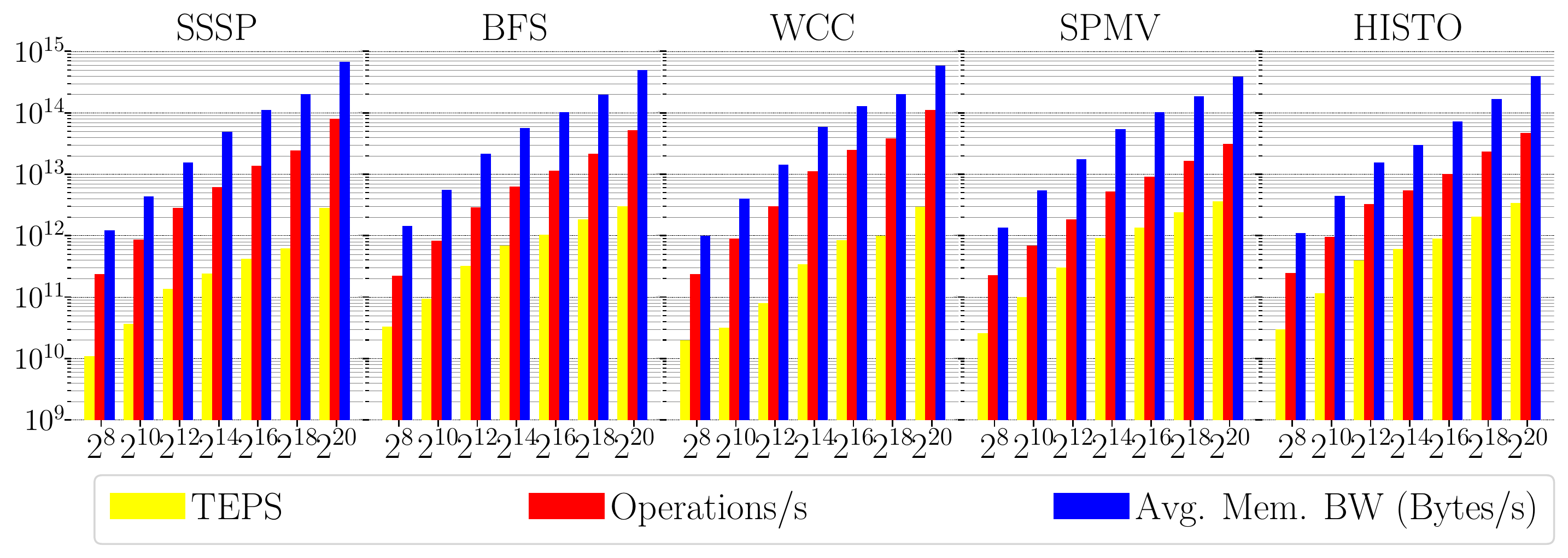}
\vspace{+1mm}
//
\hspace{-5mm} \includegraphics[width=1.05\columnwidth]{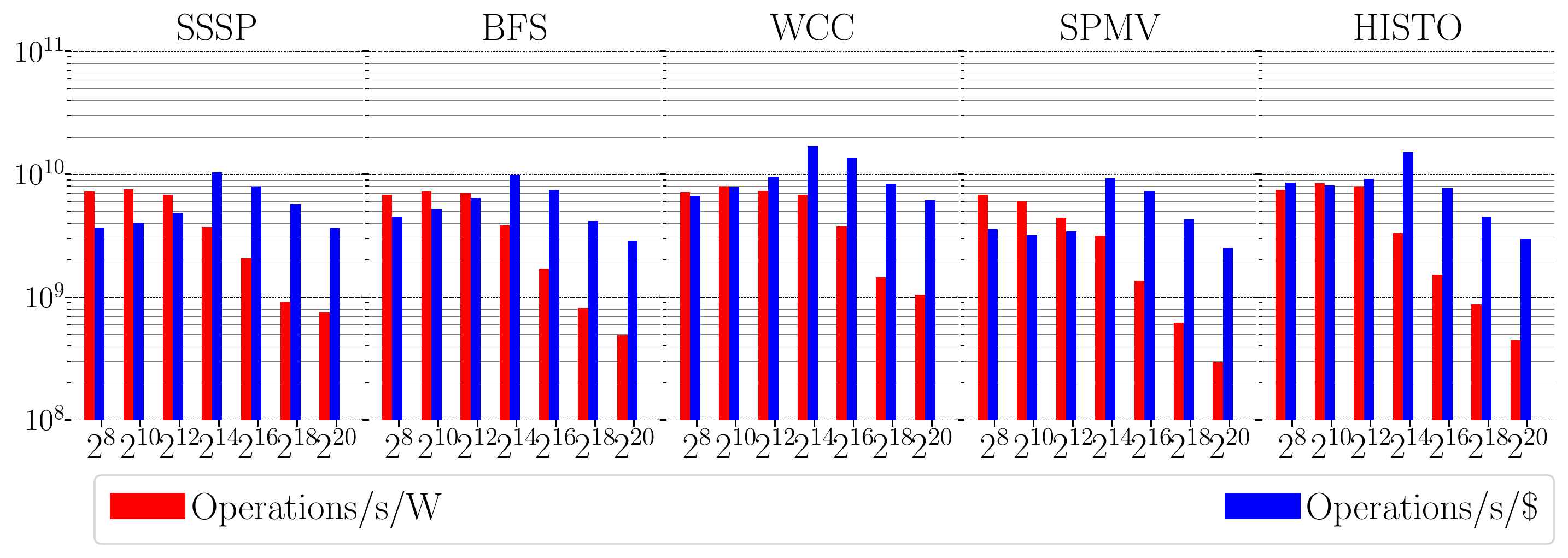}
\caption{
Throughput in operations, traversed edges per second (TEPS), and the average on-chip memory bandwidth needed to achieve that.
The X-axis is the size of the \proj grid used when analyzing strong scaling R26, ranging from 256 to over a million tiles.
The bottom plot shows throughput as a function of power and cost.
Higher is better.
}
\label{fig:scaling_million}
\end{figure}

\textbf{\textit{Throughput-per-watt likes small grids:}}
Fig.\ref{fig:scaling_million} (bottom, red bars) shows that if throughput-per-watt is the target metric, staying within the smallest configuration that fits the dataset is the correct choice.
The R26 dataset fits entirely on a 16x16 die with 8GB HBM.
Throughput-per-watt drops significantly after using $2^{12}$ tiles, i.e., a chip package of 64x64 tiles.
The inter-package links are more power-hungry than the ones inside the package, hence, the drop in efficiency.

\textbf{\textit{Throughput-per-dollar increases during superlinear performance scaling:}}
While the cost grows linearly with the number of chips, the performance grows super-linearly until the scaling step,
where the decrease in footprint-per-tile also decreases the pressure in the memory controller of the HBM. As a result,
we observe that the optimal throughput-per-dollar peaks at a 128x128 configuration (4 chip packages) (Fig.\ref{fig:scaling_million} bottom, blue bars).
At 128x128, R26 fits in SRAM, and any smaller footprint-per-tile reduces PU utilization since there are fewer parallel tasks to be executed.
This leads to 
the throughput-per-dollar to decrease beyond this point.

\textbf{\textit{Strong scaling for faster time-to-solution:}}
For purposes of this experiment, we take the extreme approach of parallelizing a dataset with $2^{26}$ vertices (and $\sim$$2^{28}$ edges) across $2^{20}$ tiles. This shows that \proj is suited for strong scaling, although it is not the most efficient way to use it.

\textbf{\textit{Petabyte/s of memory bandwidth:}}
As mentioned earlier, data-structure traversal has a high memory-to-compute ratio.
Fig.\ref{fig:scaling_million} demonstrates how much memory bandwidth is required to maintain a high target throughput.
For the 1-million-tile configuration, SPMV reads, on average, almost half a PB/s from their local memories.
At peak, SPMV reads 2.2 PB/s to perform 415 Teraops/s, of which 41 Teraops/s are dedicated to the multiplication of non-zero matrix elements.
This configuration uses 256 chip packages and draws 12KW of power on average and 40KW at its peak--- power density stays within the tens of mW/mm$^2$ suitable for air cooling.

\textbf{\textit{Comparing with the state-of-the-art:}}
Top two performances for BFS on \textbf{R26} in Graph500 list are Tianhe Exa-node (Prototype@GraphV)~\cite{graph500} and an undisclosed \textit{High Throughput Computer} with four NVidia V100-SXM2, delivering 
884 and 392 GTEPS, respectively. 
For R26, our work performs 1826 GTEPS on a 512x512 grid (128 chips) and 3021 GTEPS on a 1024x1024 grid (256 chips).
Bringing the dataset onto the chip would take $<0.1$ms, given the assumed chip I/O and the number of chips.
For smaller datasets like \textbf{R22}, best performing prior work is unlisted on Graph500 and demonstrates up to 70 GTEPS~\cite{simula_graphcore_bfs} running the Enterprise~\cite{liu2015enterprise} and Gunrock codes~\cite{wang2016gunrock} on a 32GB NVIDIA Tesla V100-SXM3.
For R22, the 64x64 configuration (single-chip) evaluated in Fig.\ref{fig:queue_characterization} achieves 362 GTEPS (5.2$\times$ improvement).
Note that this GPU runs at 1.6 GHz, while our system assumes 1 Ghz.

\section{Conclusion}\label{sec:conclusions}


This paper demonstrates that \proj scales performance for sparse-data applications even when parallelized to an unprecedented degree, such as processing $2^{26}$ vertices across $2^{20}$ PUs.
This degree of parallelization is 100$\times$ greater than the previous largest one for this problem size. Moreover, our BFS throughput is higher than the top entries of the Graph500 list.

We achieve this by introducing proxy regions and selective cascading.
Our proxy approach mitigates the work imbalance arising from the massive parallelization of sparse datasets by allowing proxy tiles to perform tasks (using copies of the data) on behalf of the data owner.
Data updates are propagated seamlessly and opportunistically (improving NoC utilization) thanks to the write-back proxy caches and selective cascading.

We put much effort into making a chiplet-based design that is highly configurable,
not only so that it could potentially be applied to other domains, but also within the same domain, to optimize the chip at packaging time for different target metrics. 
Our evaluation shows that packaging the same \device chiplets with or without on-chip DRAM leads to different throughput and energy-efficiency per-dollar results and different optimal scaling steps for throughput-per-watt.

\bibliographystyle{plain}
\bibliography{refs}

\end{document}